\documentclass[lettersize,journal]{IEEEtran}
\usepackage{amsmath,amsfonts}
\usepackage[noend]{algorithmic}
\usepackage{algorithm}
\usepackage{array}
\usepackage{textcomp}
\usepackage{stfloats}
\usepackage{float}
\usepackage{url}
\usepackage{verbatim}
\usepackage{graphicx}
\usepackage{cite}
\usepackage{tabularx}
\usepackage{array}
\usepackage{booktabs}
\usepackage{makecell}
\usepackage{caption}
\usepackage{subcaption}
\usepackage{xcolor}
\usepackage{wrapfig}
\usepackage{multirow}
\usepackage{bbding}
\newcommand{\ourwork}{Let2Gain}
\newcommand{\cmark}{\textcolor{green!60!black}{\ensuremath{\pmb{\checkmark}}}}
\newcommand{\xmark}{\textcolor{red!80!black}{\XSolidBrush}}
\begin{document}

\title{Let2Gain: Efficient LLM Inference on 2.5D Chiplet Architecture using 2T Gain-Cell Memory}

\author{%
Arkapravo~Ghosh,
XinXin~Wang,
Shuhan Liu,
Varun~Manjunath,
H.-S.~Philip~Wong,
and Priyadarshini~Panda%
\thanks{Arkapravo Ghosh, Varun Manjunath, and Priyadarshini Panda are with the Ming Hsieh Department of Electrical and Computer Engineering, University of Southern California, Los Angeles, CA 90089 USA.}%
\thanks{Shuhan Liu is with the Department of Electrical and Computer Engineering, Yale University, USA.}
\thanks{XinXin Wang and H.-S. Philip Wong are with the Department of Electrical Engineering, Stanford University, Stanford, CA 94305 USA.}%
\thanks{This work was supported in part by CoCoSys, a JUMP2.0 center sponsored by DARPA and SRC, the National Science Foundation (CAREER Award, Grant \#2312366, Grant \#2318152), the DARPA Young Faculty Award, the DoE MMICC center SEA-CROGS (Award \#DE-SC0023198) and the Global Industrial Technology Cooperation Center(GITCC) program.}
}




\maketitle

\begin{abstract}

As computation and memory demands for AI workloads grow, 2.5D accelerators offer a scalable alternative to low-yield monolithic dies. While prior work has focused on reducing data communication, on-chip SRAM dominates energy and area costs during LLM inference. 2T GainCell (GC) memories are a promising alternative due to higher density and lower-leakage than SRAM. However, their limited retention and read bandwidth have prevented their use as general-purpose cache. No prior work has explored LLM inference mapping on 2.5D accelerators while replacing the SRAM cache hierarchy with such energy-efficient but short-retention, slower memories. We present Let2Gain, an end-to-end framework evaluating 2T GainCells as on-chip cache in 2.5D LLM accelerators. Using a GainCell-aware scheduler/mapper along with data-lifetime and thermal simulation-guided refresh energy evaluation, Let2Gain shows that SRAM can be fully replaced by GainCells if occasional throughput loss is acceptable. Across diverse LLMs and serving workloads, we find that for moderately sized caches, nearly all data tiles fit within retention limits, yielding $<1\%$ refresh energy overhead. A pure GainCell cache hierarchy reduces inference energy by up to $2.15\times$, improves efficiency (tokens/s/(W-mm$^2$)) by up to $2.83\times$---though at the cost of up to $1.21\times$ slower inference than SRAM-based designs. Design space exploration (DSE) further finds that throughput-prioritized designs favor a hybrid GC-SRAM split, achieving up to $2.2\times$ efficiency at iso-runtime with SRAM, while energy-optimal designs converge on a pure GainCell stack with $\geq 2\times$ inference energy reduction.
\end{abstract}


\section{Introduction}

Large Language Models (LLMs) now power diverse applications, from conversational assistants to code generation and AI agents, motivating hardware tailored to their inference characteristics~\cite{patel2024splitwise,lin2024parrot,yu2024cambriconllm}. Under practical serving scenarios \cite{agrawal2024sarathi}, these workloads employ multi-batch inference, which sharply raises the compute and memory demands. Hence, advanced nodes have enabled larger monolithic accelerators with tens of billions of transistors~\cite{vasiljevic2021compute,knowles2021graphcore,jiao20207}, but further scaling is constrained by the reticle limit and declining yield at large die sizes~\cite{kalka2017mcmgpu,tan2021nnbaton}. Thus, partitioning accelerators into smaller chiplets offers a scalable, yield-efficient alternative~\cite{tan2021nnbaton,cai2024gemini}.

However, the need for specialized packaging using organic substrates or silicon interposers \cite{cai2024gemini} and additional PHY modules for the die-to-die (D2D) communication increases their manufacturing cost and power consumption. Hence, previous works \cite{cai2023inter, cai2024gemini, gao2019tangram, kwon2021heterogeneous, shao2019simba, das2023moham, odema2023interlayer} have focused on efficient scheduling and mapping of AI workloads on such tiled accelerators using inter-layer scheduling (ILS) and DRAM interleaving strategies to reduce the redundant activation access through D2D and the PCIe links between consecutive DNN layers.

\begin{figure}[h]
    \centering
    \includegraphics[width=\linewidth]{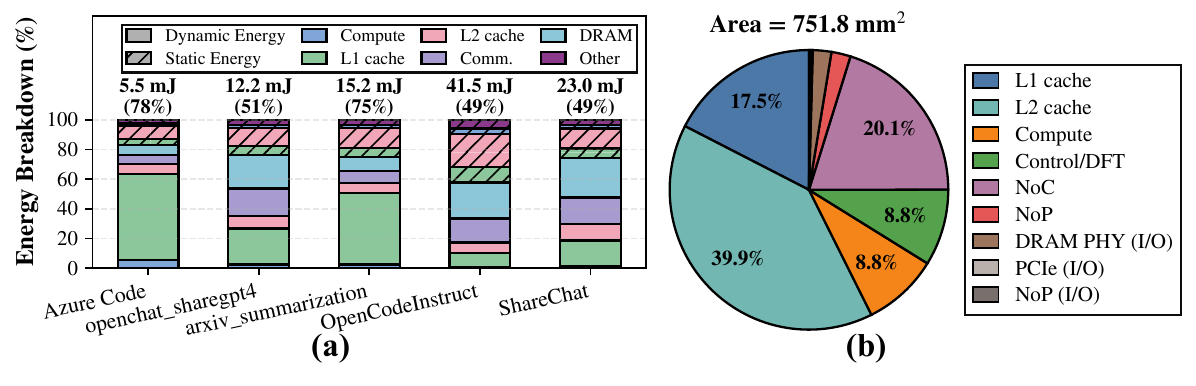}
    \caption{(a) Energy breakdown (per-token) for end-to-end inference across different LLM tasks on a Simba-\cite{shao2019simba} like 2.5D accelerator$^*$ in TSMC 40nm at a batch size of 32 for Llama-3 8B W8A8KV8. SRAM energy fraction is annotated within (). (b) Area breakdown for the 2.5D accelerator.  NoC: Network-on-chip, NoP: Network-on-Package (D2D links).}
    \label{fig:sram_overhead}
    \vspace{2mm}
    \parbox{\linewidth}{\scriptsize
*Accelerator configuration defined in Table \ref{tab:exp_setup_hw}.
}
\end{figure}

\begin{figure}[h]
    \centering
    \includegraphics[width=\linewidth]{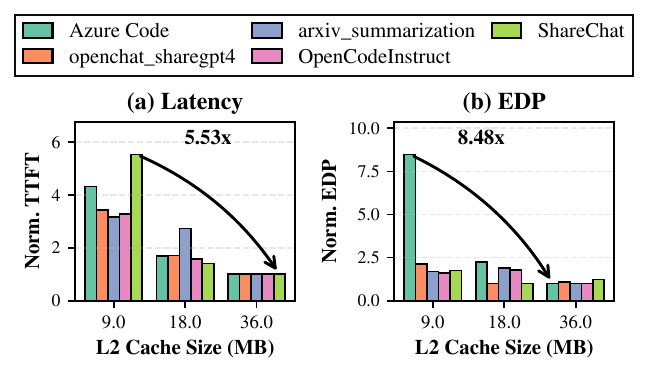}
    \caption{Impact of L2 cache size on (a) latency and (b) energy-delay product (EDP) across different tasks, in TSMC 40nm at a batch size of 32.}
    \label{fig:cache_sweep}
\end{figure}

Despite these architectural improvements, the on-chip memory hierarchy, dominated by SRAM, remains the major source of energy and area overhead. As shown in Fig.\ref{fig:sram_overhead}, the on-chip memory access of $L_1$ cache and the leakage (or static) energy of the $L_2$ cache dominate the energy consumption, accounting for \textbf{49-78\%} of the total inference energy. Fig.~\ref{fig:sram_overhead}(b) shows SRAM dominates on-package area (\textbf{57.4\%}), making it the primary hardware bottleneck. 
Further, increasing the aggregate L2 cache size for the 2.5D chiplet system in Fig.\ref{fig:sram_overhead} improves the time-to-first token (TTFT) and the end-to-end inference energy-delay product (EDP), as shown in Fig.\ref{fig:cache_sweep}. However, the lower density and high leakage power of SRAM make it prohibitive to deploy such large on-chip caches under practical power, area and thermal constraints. 

Emerging high-density, short-retention (SR) memories such as 2T GainCells (GCs)~\cite{liu2024design} offer a promising alternative to SRAM. Their compact 2T bit-cells substantially reduce area and energy and cut leakage by an order of magnitude, enabling larger caches within the same power/thermal budget~\cite{wang2025opengcram}. However, two limitations hinder GC adoption for on-chip caches: (i) limited retention, which requires periodic refreshes or costly cache write-backs to off-chip DRAM, and (ii) limited read/write bandwidth over SRAM, limiting  throughput in cache-bound systems. Their retention period also drops exponentially as temperature rises \cite{chun20113t}, a possible concern for serving systems. 

Although limited prior work~\cite{tu2018rana, xia2025kelle, li2025gainsight} explores short-retention memories for specific applications, no framework exists to evaluate the energy--latency--area (ELA) and thermal implications of 2T GCs as \textbf{general-purpose cache} in 2.5D accelerators while modeling the impact of their slower speed on the workload mapping. Table \ref{tab:comparison} qualitatively compares \ourwork{} with the existing framworks. There is an evident gap in the evaluation fidelity for replacing SRAMs with GCs, ignoring potential hybrid SRAM-GC cache for throughput-centric LLM serving.

\begin{table}[h]
\centering
\caption{Comparison with prior work.}
\label{tab:comparison}
\setlength{\tabcolsep}{3pt}
\renewcommand{\arraystretch}{1.15}
\resizebox{\columnwidth}{!}{%
\begin{tabular}{lcccccc}
\toprule
\textbf{Work} &
\makecell{\textbf{SR-cache}} &
\makecell{\textbf{Cache}\\\textbf{speed}} &
\textbf{2.5D} &
\textbf{ILS} &
\makecell{\textbf{Thermal-}\\\textbf{aware}} &
\makecell{\textbf{Serving-opt.}\\\textbf{hybrid cache}} \\
\midrule
HISIM \cite{wang2025hisim} & \xmark & \xmark & \xmark & \xmark & \cmark & \xmark \\
GEMINI \cite{cai2024gemini} & \xmark & \cmark & \cmark & \cmark & \xmark & \xmark \\
GainSight \cite{li2025gainsight} & \cmark & \xmark & \xmark & \xmark & \xmark & \xmark \textsuperscript{a} \\
RANA \cite{tu2018rana} & \cmark & \xmark & \xmark & \xmark & \xmark & \xmark \\
Kelle \cite{xia2025kelle} & \cmark\textsuperscript{b} & \cmark & \xmark & \xmark & \xmark & \xmark \\
\textbf{\ourwork{} (Ours)} & \cmark & \cmark & \cmark & \cmark & \cmark & \cmark \\
\bottomrule
\end{tabular}%
}
\vspace{2pt}
\begin{minipage}{\columnwidth}
\footnotesize
\textsuperscript{a}Ignores lower cache speed for SR-memories.
\textsuperscript{b}Key-value (KV) cache only; not for general-purpose cache.
\end{minipage}
\end{table}

\begin{figure}[h]
    \centering
    \includegraphics[width=\linewidth]{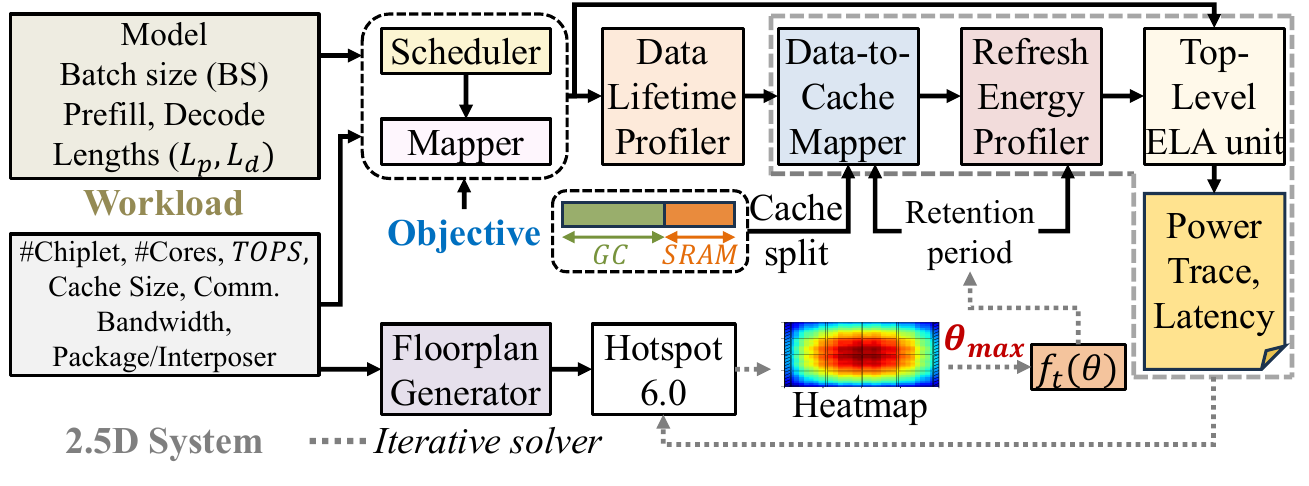}
    \caption{Overview of Let2Gain framework.}
    \label{fig:overview}
\end{figure}

 To this end, we propose \emph{Let2Gain}, an end-to-end framework for evaluating 2T GainCells as on-chip cache for LLM inference in 2.5D accelerators. As shown in Fig.~\ref{fig:overview}, Let2Gain's \emph{Mapper and Scheduler} engine (Sec.~\ref{sec:sched}) takes the LLM workload and 2.5D system description to find an ILS and layer-to-core mapping based on the performance objective. The \emph{Data Lifetime Profiler} (Sec.~\ref{sec:lifetime}) uses this mapping to determine operand lifetimes across cache levels, which the \emph{Data-to-Cache Mapper} (Sec.~\ref{sec:data-to-cache}) combines with the target GainCell--SRAM split to assign data tiles to cache addresses for minimizing the refresh energy. The \emph{Refresh Energy Profiler} (Sec.~\ref{sec:refresh}) computes this energy, which along with the Scheduler and Mapper output, drives the \emph{Top-Level ELA unit} (Sec.~\ref{sec:ela}) to produce power and latency traces. Separately, the \emph{Floorplan Generator} (Sec.~\ref{sec:floorplan}) uses the system specifications to generate a floorplan, fed to HotSpot 6.0~\cite{zhang2015hotspot} for a thermal heatmap. Since GainCell retention depends on temperature ($f_t(\theta)$) and in turn affects ELA as well as temperature, Let2Gain uses an \emph{Iterative Solver} (Sec.~\ref{sec:iter_solver}) to reach the steady-state metrics.

The key contributions of our work can be summarized as follows:
\begin{enumerate}
    \item We propose \emph{Let2Gain}, the first end-to-end framework to
    evaluate 2T GainCells as \emph{general-purpose} cache in 2.5D systems,
    jointly modeling energy, latency, area, and temperature.

    \item We demonstrate that the data lifetimes in $L_1$ and $L_2$ cache are well within the retention limits of Si-Si and OS-Si GC respectively, resulting in $<1\%$ refresh energy overhead after replacing SRAM fully with GainCell cache on a Simba\cite{shao2019simba}-like accelerator.

    \item Across 25 model–workload pairs, a 2T GainCell-based cache hierarchy improves inference energy and hardware efficiency by up to $2.15\times$ and $2.83\times$, respectively, and lowers peak temperature by up to $17^{\circ}$C, at the cost of at most $1.21\times$ slower inference.

    \item We perform comprehensive DSE to identify the optimal hybrid GC-SRAM cache composition under different optimization targets , which enables iso-runtime serving with up to $2.2\times$ higher efficiency.

\end{enumerate}
To the best of our knowledge, Let2Gain is the first work to comprehensively
study the energy--retention--bandwidth trade-offs of 2T GainCells, \textbf{showing that
the SRAM stack can be replaced entirely at little or no cost to
inference speed.}
\section{Related Works}
\label{sec:related_work}
\subsection{2.5D Accelerator Simulators} HISIM \cite{wang2025hisim} maps the entire model to the chiplets, which is impractical for billion-parameter LLMs. Simba \cite{shao2019simba}, Gemini~\cite{cai2024gemini}, NN-Baton~\cite{tan2021nnbaton}, M2M \cite{zhang2024m2m}, MoHaM \cite{das2023moham}, \cite{odema2023interlayer} and \cite{huang2026scope} jointly explore chiplet architecture and spatio-temporal mapping, distributing consecutive DNN layers across chiplets to improve data reuse and reduce inter-chiplet communication. TESA \cite{shukla2023temperature} and ThermoDSE \cite{peng2026thermodse} perform temperature-aware design and workload mapping for chiplet-based accelerators. \textit{None of these works, however, explore replacing the SRAM memory stack, fully or partially to achieve hardware efficiency.}

\subsection{Short-retention (SR) Memory for On-Chip Cache}
RANA~\cite{tu2018rana} uses embedded DRAM (eDRAM) with retention-aware scheduling in a convolution neural network (CNN) training accelerator while GainSight~\cite{li2025gainsight} proposes a hybrid GC-SRAM cache---both targeting monolithic, non-tiled  accelerators. Kelle~\cite{xia2025kelle} applies eDRAM to KV caching for edge LLM serving, using attention-based eviction to cut refresh energy, but adds a dedicated KV-cache hierarchy rather than a general-purpose cache. \textit{However, none of these works consider how the lower read bandwidth of such SR memories affects system throughput.}


\section{Background \& Motivation}
\label{sec:bgnd_motive}
\subsection{2.5D Chiplet-based Tiled-Accelerators}
\label{sec:bgnd_chiplet}
\begin{figure}[h]
    \centering
    \includegraphics[width=\linewidth]{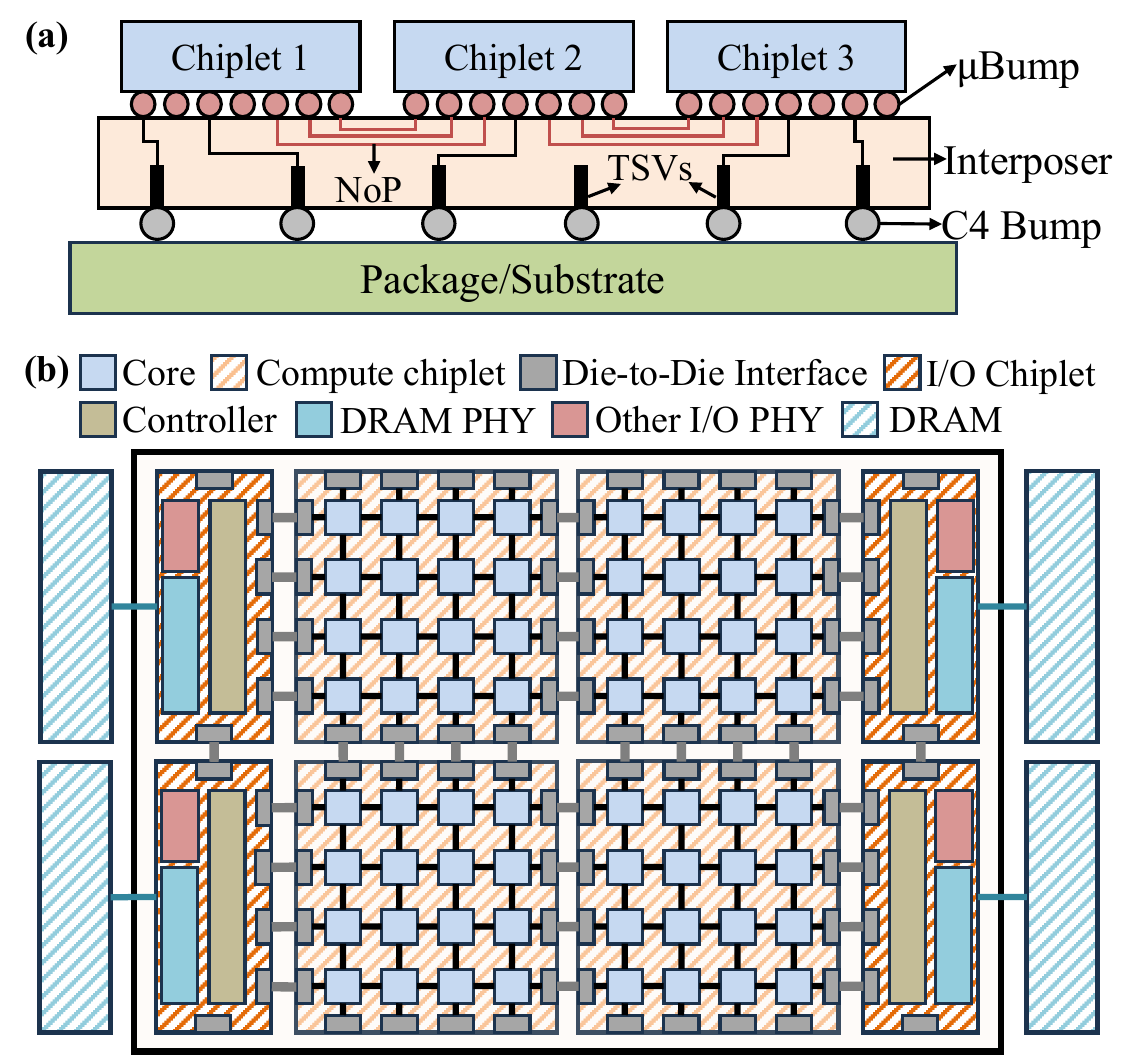}
    \caption{(a) Cross-sectional and (b) top view of a Simba\cite{shao2019simba}-like 2.5D chiplet-based deep neural network (DNN) accelerator. \textit{Network On Package} (NoP) is used to communicate between dies/chiplets. \textit{Through-silicon vias} (TSVs) are used for connecting the interposer with the substrate.}
    \label{fig:chiplet}
\end{figure}
As shown in Fig.~\ref{fig:chiplet}(a), \textit{2.5D chiplet integration} places multiple smaller dies side-by-side on a silicon or organic interposer. Chiplets communicate over a Network-on-Package (NoP) on the interposer via dense $\mu$-bumps, while through-silicon vias (TSVs) and C4 bumps connect the interposer to the substrate~\cite{park2020design}. Fig.~\ref{fig:chiplet}(b) shows the top view of the 2.5D multi-chiplet accelerator comprising of multi-core \textit{compute chiplets} with \textit{die-to-die (D2D) interfaces} (NoP I/O in Fig.\ref{fig:sram_overhead}(b)), whose cores are linked by a mesh network-on-chip (NoC), and \textit{I/O chiplets} that connect them to the DRAM channels.

\begin{wrapfigure}{r}{0.5\linewidth}
    \centering
    \vspace{-10pt}
    \includegraphics[width=\linewidth]{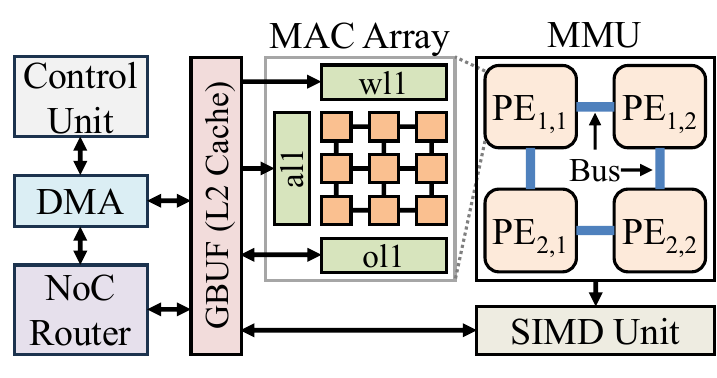}
    \caption{Architecture of the core shown in Fig.~\ref{fig:chiplet}(b).}
    \label{fig:core}
\end{wrapfigure}

Fig.\ref{fig:core} shows the architecture of the core shown in Fig.\ref{fig:chiplet}. Each core consists of a matrix multiplication unit (MMU) containing processing engine (PE) array connected by bus. Each PE consists of a multiply-and-accumulate (MAC) array comprising of its local L1 cache for input activations (al1), weights (wl1) and outputs (ol1). The PEs share a global buffer (GBUF, L2 cache) which feeds data to and from the L1 caches.
 A dedicated single instruction multiple data (SIMD) unit is present for executing normalization, activation and softmax operations. The communication unit comprises of DMA and router which handle data transfer to other cores and DRAMs, while the control unit manages computation based on statically-compiled instructions and coordinates data exchange with external cores and DRAMs.

\subsection{2T GainCell Memory}
\label{sec:bgnd_gc}
A 2T GainCell (2T-GC) is a capacitorless, fully logic-compatible eDRAM cell that stores data on a transistor gate, with separate write and read transistors~\cite{chun2012logic}. The write transistor sets the charge on the storage node, and a read senses the resulting gate-voltage-dependent current~\cite{somasekhar20092}. With fewer transistors and no cross-coupled inverters, GCs are smaller and leak less than 6T SRAM, but their stored charge decays, requiring periodic refresh~\cite{somasekhar20092,chun2010asymmetric}.
2T-GCs are further classified by transistor material. \emph{2T Si-Si GCs} use silicon for both transistors, offering standard logic-flow compatibility and fast reads. \emph{2T OS-Si GCs} use an oxide-semiconductor (e.g., IGZO) write transistor, whose ultra-low off-state leakage greatly extends retention. Additionally, BEOL fabrication makes them denser than both Si-Si GCs and 6T SRAM. However, their lower carrier mobility makes their write bandwidth lower than that of both alternatives~\cite{wang2025opengcram}.
\begin{table}[h]
\centering
\caption{Comparison of 6T SRAM and 2T GainCell memory macro.}
\label{tab:memory_comparison}
\resizebox{\columnwidth}{!}{%
\begin{tabular}{c|c|c|c|c|c}
\hline
\textbf{Memory*} &
\textbf{\shortstack{Dynamic\\energy/bit(pJ)}} &
\textbf{\shortstack{Leakage\\power (mW)}} &
\textbf{\shortstack{Area/bit\\($\mu$m$^2$/bit)}} &
\textbf{\shortstack{Retention\\period ($\mu$s)}} & \textbf{\shortstack{Bandwidth\\(GB/s)}} \\
\hline

\textbf{6T SRAM} &
1.56 &
263.64 &
0.80 &
N/A &
256\\
\hline

\textbf{2T Si-Si GC} &
0.50 &
29.29 &
0.62 &
20 &
128\\
\hline

\textbf{2T OS-Si GC} &
1.30 &
\shortstack{26.37} &
0.40 &
200 &
64\\
\hline
\end{tabular}%
}
\vspace{2pt}
    \parbox{\linewidth}{\scriptsize
*Cache size = 1MB, \#banks = 32, and block size = 1024 bits.
}
\end{table}
Table~\ref{tab:memory_comparison} compares 6T SRAM and 2T GC macros in TSMC 40nm, simulated with OpenGCRAM~\cite{wang2025opengcram}. Since GCs use a single-ended read path to the storage node, they lack SRAM's differential bitline sensing, which limits their read bandwidth.\\

\noindent \textbf{Motivation:}
Fig.~\ref{fig:sram_overhead} shows that the $L_1$ and $L_2$ caches are the dynamic-energy and area bottlenecks, respectively. Hence, it motivates us to replace $L_1$ with 2T Si-Si GCs, which offer lower dynamic energy, and $L_2$ with 2T OS-Si GCs, which offer higher density. This also matches retention to data lifetime: shorter (longer)-lived $L_1 (L_2)$ data suits the shorter (longer) retention period of Si-Si (OS-Si) GC.

\section{Let2Gain Framework}
\label{sec:framework}
We present a detailed explanation of the individual components of Let2Gain depicted in Fig.\ref{fig:overview}. Table~\ref{tab:notation} summarizes the notation used in this work. Please note that the terms buffer and cache are used interchangeably in this work.
\subsection{Mapper and Scheduler}
\label{sec:sched}

\begin{table*}[h]
\centering
\caption{Notation used in this work.}
\label{tab:notation}
\begin{tabular}{@{}ll@{}}
\toprule
\textbf{Symbol} & \textbf{Meaning} \\
\midrule
\multicolumn{2}{@{}l}{\textit{Workload}} \\
$G=(V,E)$ & DNN as a DAG; $V=\{1,\dots,L\}$ layers, $(i,j)\in E$ $\Rightarrow$ layer $i$ feeds layer $j$ \\
$L$ & number of layers in the DNN \\
$\ell$ & index of a layer, $\ell\in V$ \\
$A, F, O$ & activation, filter (weight) and output operands respectively \\
$d_\ell \in \{C_\ell, K_\ell, H_\ell, W_\ell, R_\ell, S_\ell\}$ & input-channel, output-channel, output height, output width, weight height and weight width of layer $\ell$ \\
$B$ & batch size \\
\midrule
\multicolumn{2}{@{}l}{\textit{Hardware}} \\
$a\times o$ & PE-group array dimensions per core \\
$v$ & channel-parallel width of the MAC array \\
$x_c \times y_c$ & 2D core mesh within a compute die \\
$AL_1,WL_1,OL_1$ & private per-PE-group buffers for different operands \\
$L2$ & buffer shared across all PE groups on a core \\
$X$ & cache type $\in{[AL_1, WL_1, OL_1, L_2]}$ \\
$\beta_X^{tech}$ & bandwidth of buffer $X$ of memory technology $tech \in [GC, SRAM]$ \\
$S_X$ & capacity of cache $X$ \\
$Comm$ & communication components, $Comm \in [NoC, NoP, DRAM]$ \\
$X_c \times Y_c$ & 2D die mesh for the entire 2.5D system, \#Chiplets=$X_cY_c$ \\
\midrule
\multicolumn{2}{@{}l}{\textit{Mapper (Spatial Mapping)}} \\
$\sigma_\ell$ & a spatial mapping of a layer $\ell$ onto a core \\
$(d_A,d_O)$ & the two workload dimensions unrolled across the $a$- and $o$-axes in PE array\\
$\Sigma(\ell)$ & set of candidate mappings the Mapper searches over for layer $\ell$ \\
$E(\sigma_\ell),T(\sigma_\ell)$ & energy, latency of layer $\ell$ under mapping $\sigma$ \\
\midrule
\multicolumn{2}{@{}l}{\textit{Scheduler (Inter-Layer Schedule)}} \\
$\tau$ & a schedule which is a tree of cuts over the layers \\
$E_{tot}(\tau),T_{tot}(\tau)$ & total energy, latency of schedule $\tau$ \\
\midrule
\multicolumn{2}{@{}l}{\textit{Modeling GainCell}} \\
$f_X\in[0,1]$ & fraction of level $X$ cache implemented as GainCell \\
$\kappa_X$ & per-access energy scale factor of gain-cell relative to SRAM at level $X$ \\
$t_{ret}$ & Retention period (in seconds) \\
\midrule
\multicolumn{2}{@{}l}{\textit{Data Lifetime Profiler}} \\
  $\mu_0(\ell)$ & cycles for one innermost reduction pass \\
  $T_{L1}(\ell)$ & cycles to execute one full L1 tile loop \\
  $\mathrm{rel}(X)$ & workload dimensions that affect buffer $X$'s data lifetime \\
$\Lambda_X(\ell)$ & lifetime of one tile of layer $\ell$ in $X$, in cycles \\
$n_X(\ell)$ & tile-instance count of $X$ over layer $\ell$ \\
$\mathcal{P}_X$ & set of data tiles in cache level $X$ \\
$f_{clk}$ & core clock frequency \\
\bottomrule
\end{tabular}
\end{table*}

\paragraph{Workload and Hardware Representation} An $L$-layer neural network is a DAG $G=(V,E)$, where each vertex $\ell\in V$ is a layer and edge $(i,j)\in E$ means layer $j$ consumes layer $i$'s output. Layer $\ell$ is described by its input/output channels $C_\ell$/$K_\ell$, output height/width $H_\ell$/$W_\ell$, filter height/width $R_\ell$/$S_\ell$, and batch size $B$. Hardware template follows the description of Sec.\ref{sec:bgnd_chiplet} (Fig.\ref{fig:chiplet}(b), \ref{fig:core}).

\paragraph{Mapper} For each layer $\ell$, the Mapper decides how its computation is spread across PE grid of multiple cores and staged through the memory hierarchy.
  We denote this decision by the \emph{spatial mapping} $\sigma_\ell$. $\sigma_\ell$ spatially unrolls a dimension pair $(d_A,d_O)\in\{C_\ell,K_\ell,H_\ell,W_\ell,B\}^2$, $d_A\neq d_O$, across the PE grid. The remaining dimensions are iterated temporally, with tile sizes bounded by cache capacities. The energy and latency of layer $\ell$ under $\sigma_\ell$ are
  \begin{equation}
\begin{split}
  E(\sigma_\ell) = {}& E_{comp}(\sigma_\ell) + \sum_{X} E_X(\sigma_\ell) + \sum_{Comm} E_{Comm}(\sigma_\ell)
\end{split}
\label{eq:energy}
\end{equation}
\begin{equation}
\begin{split}
  T(\sigma_\ell) = {}& \max\Big(T_{comp}(\sigma_\ell),\ \max_{X} T_X(\sigma_\ell), \ \max_{Comm} T_{Comm}(\sigma_\ell)\Big)
\end{split}
\label{eq:time}
\end{equation}
  where $E_{comp}$, $T_{comp}$ denote compute energy and time; $E_X$, $T_X$ the access energy and streaming time of buffer $X$; and $E_{Comm}$, $T_{Comm}$ for $Comm\in\{\text{NoC},\text{NoP},\text{DRAM}\}$ the data-movement energy and time over the NoC, NoP and DRAM respectively. We define $\Sigma(\ell)$ as the set of all valid spatial mappings for the layer $\ell$, defined by
  \begin{equation}
\begin{split}
  \Sigma(\ell) = {}& \big\{\, (d_A,d_O,\{t_X\}) \;:\; d_A,d_O\in\{C,K,H,W,B\},\ 
  \\ & d_A\neq d_O, t_X\in\mathbb{Z}^+,\ t_X\le S_X \\ & \forall X\in\{AL_1,WL_1,OL_1,L2\} \,\big\},
\end{split}
\label{eq:sigma-space}
\end{equation}
where, $t_X$ is the tile size assigned to buffer $X$ and must be within its capacity ($S_X$). The Mapper selects the most optimal mapping for layer $\ell$ which minimizes the objective (Fig.\ref{fig:overview}).  
  \begin{equation}
  \sigma_\ell^* = \arg\min_{\sigma_\ell\in\Sigma(\ell)} \mathrm{Cost}\big(E(\sigma_\ell),T(\sigma_\ell)\big),
  \label{eq:mapper-opt} 
  \end{equation}

where, $Cost$ is the user objective, which can be energy-delay product (EDP) or a custom-defined metric. 

\paragraph{Scheduler} The Scheduler decides how the $L$ layers are organized across the accelerator cores (spatially) and time (temporally), producing the \emph{inter-layer schedule} (ILS), represented as a tree $\tau$. Each leaf of $\tau$ is a single layer, mapped by the \emph{Mapper} to a set of core clusters as described previously. Each internal node $n$ is a \emph{cut} with children $c_1,\dots,c_m$, of one of two types, defined as follows.
\begin{itemize}
\item \textbf{S (spatial)-cut} $n$'s core cluster is partitioned into $m$ disjoint subsets, one per child, and the children execute \emph{concurrently}, each on its own subset.
\item \textbf{T (temporal)-cut} $n$'s entire core cluster is given to each child in turn, and the children execute \emph{sequentially}.
\end{itemize}
The two cut types aggregate cost differently:
\begin{align}
\text{S-cut:} \quad & E_n = \textstyle\sum_{i=1}^m E_{c_i}, \qquad T_n = \max_{i=1,\dots,m} T_{c_i}, \label{eq:scut}\\
\text{T-cut:} \quad & E_n = \textstyle\sum_{i=1}^m E_{c_i}, \qquad T_n = \textstyle\sum_{i=1}^m T_{c_i}. \label{eq:tcut}
\end{align}
The leaves' costs $\big(E(\sigma_\ell)),T(\sigma_\ell)\big)$ from the Mapper (Eq.\ref{eq:energy}, \ref{eq:time}) propagate up through $\tau$ via the recursion
  \begin{equation}
  (E_n,T_n) =
  \begin{cases}
  \big(E(\sigma_{\ell}),\, T(\sigma_{\ell})\big) & \text{leaf } (\text{layer } \ell) \\[4pt]
  \Big(\sum_{i=1}^{m} E_{c_i},\ \max\limits_{i} T_{c_i}\Big) & \text{S-cut} \\[6pt]
  \Big(\sum_{i=1}^{m} E_{c_i},\ \sum_{i=1}^{m} T_{c_i}\Big) & \text{T-cut}
  \end{cases}
  \label{eq:cost-recursion}
\end{equation}

This recursively yields the schedule's total cost $\big(E_{tot}(\tau),T_{tot}(\tau)\big)$ at the root of $\tau$.
\begin{equation}
  \mathcal{T} = \Big\{\, \tau \;:\; \
  c(n)\in\{leaf,S,T\}\ \forall n,\ \ b(n)\in\mathbb{Z}^{+}\ \forall n \,\Big\}
  \label{eq:sched-space}
\end{equation}
  is the set of admissible schedules, where for a tree $\tau\in\mathcal{T}$, $c(n)$ is the cut type of internal node $n$ and $b(n)$ is the batch size assigned
  to node $n$, which is constant along a T-cut segment. The Scheduler then solves
  \begin{equation}
  \tau^* = \arg\min_{\tau\in\mathcal{T}} \mathrm{Cost}\big(E_{tot}(\tau),T_{tot}(\tau)\big),
  \label{eq:sched-opt}
  \end{equation}
  i.e.\ jointly over the tree topology, the cut-type assignment $c(\cdot)$ and the batch-size assignment $b(\cdot)$ at every internal node. To optimize the intra-layer mappings using the \textit{Mapper} for a given $\tau$, we exhaustively search all the loop unrolling strategies and choose the one that minimizes the \textit{Cost}.
The search for $\tau^*$ proceeds in two stages: an initial dynamic-programming pass constructs a starting schedule, which is then refined by simulated annealing, as detailed next.

\paragraph{Modeling GainCell} Either buffer level $X\in\{L1,L2\}$ may be built as a mix of GainCell (a fraction $f_X$ of its capacity with bandwidth $\beta_X^{GC}$) and SRAM (the remaining fraction with bandwidth $\beta_X^{SRAM}$), realized as two independently-addressed banks operating in parallel. This modifies the energy (Eq.\ref{eq:energy}) and latency (Eq.\ref{eq:time}) model for the \emph{Mapper}. For layer $\ell$, latency is set by whichever of the two concurrent banks is slower,
\begin{equation}
T_X^{'}(\sigma_\ell) = \max\!\left(\frac{f_X}{\beta_X^{GC}},\ \frac{1-f_X}{\beta_X^{SRAM}}\right)\times(\text{data moved through }X),
\label{eq:gc-time}
\end{equation}
while energy is the access-weighted mix of both banks' per-access costs, with a level-specific GC energy scale factor $\kappa_X$.
\begin{equation}
E_X^{'}(\sigma_\ell) = E_X(\sigma_\ell)\left(\frac{f_X}{\kappa_X} + (1-f_X)\right).
\label{eq:gc-energy}
\end{equation}
$f_X=0$ recovers the SRAM-only energy and latency (Eq.\ref{eq:energy}, \ref{eq:time}). Both Eq. \ref{eq:gc-time}, \ref{eq:gc-energy} substitute $T_X(\sigma_\ell),E_X(\sigma_\ell)$ in the Mapper's search which consequently affects the Scheduler's search Eq.\ref{eq:sched-opt} as well. Note that Eq.\ref{eq:gc-time} affects the latency (Eq.\ref{eq:time}) only when cache access becomes a bottleneck due to the $\max$ operation.

\paragraph{Search Algorithm} The search for $\tau^*$ in \eqref{eq:sched-opt} proceeds in two stages. First, a dynamic-programming pass treats the network as a
  linear chain and partitions it into contiguous segments $[j,i]$, each executed as one unit; with $C(j,i)$ the best cost of segment $[j,i]$ (minimized over 
  its batch size),
  \begin{equation}
  \mathrm{dp}[i] = \min_{0\le j\le i}\big(\mathrm{dp}[j-1] + C(j,i)\big), \qquad \mathrm{dp}[-1]\equiv 0,
  \label{eq:dp}
  \end{equation}
  gives the best segmentation, and backtracking yields an initial schedule $\tau_0$. Second, simulated annealing refines $\tau_0$ by repeatedly proposing a series of 
  local restructuring moves used in \cite{cai2024gemini} to obtain a candidate $\tau'$, accepted with probability
  \begin{equation}
  \Pr[\text{accept }\tau'] =
  \begin{cases}
  1 & \mathrm{Cost}(\tau')\le\mathrm{Cost}(\tau) \\
  \exp(-\Delta/Temp) & \text{otherwise},
  \end{cases}
  \label{eq:sa-accept}
\end{equation}
  where $\Delta$ is the relative cost increase of $\tau'$ over $\tau$. The temperature $Temp$ is high at the initial iterations for escaping local minima and is lowered over the search. The best schedule 
  found is returned as $\tau^*$.

\subsection{Data Lifetime Profiler}
\label{sec:lifetime}
\ourwork{} uses a Polar dataflow~\cite{cai2023inter} for each core (Fig.~\ref{fig:chiplet}(b)), with the loop order shown in Algorithm~\ref{alg:looporder}. Dimension $d$ has tile counts $N_d^{(1)}$ at the private $L_1$ buffers ($AL_1$, $WL_1$, $OL_1$) and $N_d^{(2)}$ at the shared $L_2$ buffer, covering its full range in $N_d^{(1)}\times N_d^{(2)}$ tiles. Note that these tile count ranges for each layer are already obtained from the Mapper and Scheduler (Sec.\ref{sec:sched}) using Eq.\ref{eq:mapper-opt} and \ref{eq:sched-opt}.
The outer loops ($k_2$, $h_2$, $w_2$) iterate over $L_2$ tiles, each staged from DRAM or the NoC. Nested inside, a batch loop iterates over all $B$ elements, followed by the inner loops ($k_1$, $h_1$, $w_1$), which stage $AL_1$, $WL_1$, and $OL_1$ tiles from the current $L_2$ tile. The innermost step is the PE array's MAC reduction over $C$, $R$, and $S$ dimensions.
\begin{algorithm}
  \caption{Loop order for executing a layer}
  \label{alg:looporder}
  \begin{algorithmic}[1]
  \FOR{$k_2=1$ \textbf{to} $N_{K_\ell}^{(2)}$}
    \FOR{$h_2=1$ \textbf{to} $N_{H_\ell}^{(2)}$}
      \FOR{$w_2=1$ \textbf{to} $N_{W_\ell}^{(2)}$}
        \STATE stage $L2$ tile
        \FOR{$b=1$ \textbf{to} $B$}
          \FOR{$k_1=1$ \textbf{to} $N_{K_\ell}^{(1)}$}
            \FOR{$h_1=1$ \textbf{to} $N_{H_\ell}^{(1)}$}
              \FOR{$w_1=1$ \textbf{to} $N_{W_\ell}^{(1)}$}
                \STATE stage $AL_1,WL_1,OL_1$ tiles
                \STATE $O_{k,h,w} \mathrel{+}= \displaystyle\sum_{c=1}^{C_\ell}\sum_{r=1}^{R_\ell}\sum_{s=1}^{S_\ell} A_{c,\,h+r,\,w+s}\cdot F_{k,c,r,s}$
              \ENDFOR
            \ENDFOR
          \ENDFOR
        \ENDFOR
      \ENDFOR
    \ENDFOR
  \ENDFOR
  \end{algorithmic}
\end{algorithm}

\noindent \textbf{Lifetime derivation for $L_1$:}
The innermost step reduces $C_\ell$ in $N_C^{(0)}(\ell)=\Big\lceil \frac{C_\ell}{v}\Big\rceil$ sequential groups of $v$ channels:
\begin{equation}
\mu_0(\ell) = N_C^{(0)}(\ell)\,R_\ell\,S_\ell \quad (cycles)
\label{eq:tau0}
\end{equation}

Each buffer $X$ has a set of relevant dimensions $\mathrm{rel}(X)\subseteq\{K_\ell,H_\ell,W_\ell,B\}$: $\mathrm{rel}(AL_1)=\{H_\ell,W_\ell,B\}$, $\mathrm{rel}(WL_1)=\{K_\ell\}$, and $\mathrm{rel}(OL_1)=\{K_\ell,H_\ell,W_\ell,B\}$. A tile in $X$ is reused until a dimension in $\mathrm{rel}(X)$ advances, at which point it is replaced. Hence, the lifetime of an $L_1$ tile in cycles is the product of the $L_1$-level tile counts of dimensions \emph{not} in $\mathrm{rel}(X)$ multiplied by $\mu_0(\ell)$ (Eq.~\ref{eq:tau0}):
\begin{align}
\Lambda_{AL_1}(\ell) &= N_K^{(1)}\cdot\mu_0(\ell), \\
\Lambda_{WL_1}(\ell) &= N_H^{(1)}N_W^{(1)}\cdot\mu_0(\ell), \\
\Lambda_{OL_1}(\ell) &= \mu_0(\ell).
\end{align}
\noindent \textbf{Lifetime derivation for $L_2$:}
The total number of tiles for the layer $\ell$ resident at each cache $X$ is the
product of the L1- and L2-level tile counts of the dimensions
\emph{in} $\mathrm{rel}(X)$:
\begin{equation}
\begin{aligned}
n_{AL_1}(\ell) &= N_{H,\ell}\,N_{W,\ell}\,B, \\
n_{WL_1}(\ell) &= N_{K,\ell}, \\
n_{OL_1}(\ell) &= N_{K,\ell}\,N_{H,\ell}\,N_{W,\ell}\,B,
\end{aligned}
\end{equation}
with $N_{d,\ell}=N_{d,\ell}^{(1)}N_{d,\ell}^{(2)}$ tiles along dimension $d$.
The time (in cycles) taken to execute the L1 loops (Lines 6-10 in Algorithm \ref{alg:looporder}) is given by
\begin{equation}
T_{L1}(\ell) = \mu_0(\ell)\,N_{K_\ell}^{(1)}N_{H_\ell}^{(1)}N_{W_\ell}^{(1)}.
\label{eq:tl1}
\end{equation}
Thus, a tile held in the $L_2$ buffer remains resident while the loop indices $h_2,w_2,b$ sweep their full ranges beneath a fixed $k_2$. The data lifetimes and tile counts for activation ($A$) and filter ($F$) at the $L_2$ cache are thus, given by
  \begin{align}
  \Lambda_{A,L_2}(\ell) &= \Lambda_{F,L_2}(\ell)= N_{H_\ell}^{(2)} N_{W_\ell}^{(2)} B \cdot T_{L1}(\ell), \\
  n_{A, L_2}(\ell) &= n_{F, L_2}(\ell)= N_{K_\ell}^{(2)}.
\end{align}
  The output tile, produced once per combination of $k_2,h_2,w_2,b$, has lifetime and tile count given by
  \begin{align}
  \Lambda_{O, L_2}(\ell) &= T_{L1}(\ell'), \\
  n_{O, L_2}(\ell) &= N_{K_\ell}^{(2)} N_{H_\ell}^{(2)} N_{W_\ell}^{(2)} B,
\end{align}
where $\ell'$ is the layer directly consuming layer $\ell$'s output.

\subsection{Data-to-Cache Mapper}
\label{sec:data-to-cache}
Each cache level $X\in\{L_1,L_2\}$ is split into parallel GC and SRAM caches, holding fractions $f_X$ and $1-f_X$ of its capacity. All $L_1$ caches share the same $f_{L_1}$. The \emph{Data-to-Cache Mapper} (Algorithm~\ref{alg:cachemapper}) assigns each tile instance to a physical cache using the lifetimes from the Data Lifetime Profiler (Sec.~\ref{sec:lifetime}). Let $\mathcal{P}_X$ be the set of tiles at level $X$, where tile $r$ has $n_r$ instances and lifetime $\lambda_r=\Lambda_r/f_{clk}$ seconds. Let $N_X=\sum_{r\in\mathcal{P}_X}n_r$. The mapper then sorts $\mathcal{P}_X$ by ascending lifetime, fills the GC cache up to its budget of $f_XN_X$ tiles, and places the rest in SRAM. This way, refresh energy is minimized by assigning the shortest-lived data to GC (discussed next).

\begin{algorithm}[h]
  \caption{Data-to-Cache Mapper}
  \label{alg:cachemapper}
  \begin{algorithmic}[1]
  \STATE $X\in\{AL_1,WL_1,OL_1,L2\}$
  \STATE sort $\mathcal{P}_X$ ascending by $\lambda_r$, giving $r_1,\dots,r_{|\mathcal{P}_X|}$
  \STATE $\mathit{budget}\gets f_X N_X$
  \FOR{$i=1$ \textbf{to} $|\mathcal{P}_X|$}
    \STATE $a_i \gets \min(n_{r_i},\,\max(0,\mathit{budget}))$ 
    \STATE place $a_i$ tiles of $r_i$ in the GainCell cache
    \STATE place $n_{r_i}-a_i$ tiles of $r_i$ in the SRAM cache
    \STATE $\mathit{budget}\gets \mathit{budget}-a_i$
  \ENDFOR
  \end{algorithmic}
  \end{algorithm}

\subsection{Refresh Energy Profiler}
\label{sec:refresh}
For each level $X\in\{L1,L2\}$, a tile $r\in\mathcal{P}_X$ placed in the gain-cell bank retains its data for at most $t_{ret,X}$ before it must be rewritten. If its lifetime $\lambda_r$ exceeds $t_{ret,X}$, it is refreshed periodically for as long as it remains resident. The number of refreshes needed beyond the initial write is
\begin{equation}
\rho_r =
\begin{cases}
0, & \lambda_r \le t_{ret,X} \\[2pt]
\big\lceil \lambda_r / t_{ret,X} \big\rceil - 1, & \lambda_r > t_{ret,X}.
\end{cases}
\label{eq:refresh-count}
\end{equation}

Each tile refresh requires reading and rewriting it in full. The energy this costs per byte is the summation of GainCell read and write energies per byte at level $X$, $e_X^{GC}=e_{X,GC}^{read}+e_{X,GC}^{write}$. Let $\mathcal{G}_X={r\in\mathcal{P}_X}$ denote the tiles at level $X$ with at least one instance placed in the GainCell cache. Summing the refresh cost over 
  every such tile placed in the GainCell cache gives the total refresh energy at level $X$:
  \begin{equation}
  E_X^{refresh} = e_X^{GC}\sum_{r\in\mathcal{G}_X} a_r\rho_rb_r,
  \label{eq:refresh-energy}
  \end{equation}
where ${b_r}$ indicates the number of bytes in tile $r$. 
$E_X^{refresh}=0$ whenever every gain-cell-resident tile satisfies $\lambda_r\le t_{ret,X}$, and increases only as tiles with $\lambda_r>t_{ret,X}$ are admitted into the GainCell cache.

\subsection{Top-Level ELA unit}
\label{sec:ela}
The end-to-end energy ($E_{sys}$) and latency ($L_{sys}$) formulations are derived as follows.
\begin{equation}
\begin{split}
  E_{sys}(\tau) = E_{tot}(\tau)+\sum_{X}E_X^{refresh}
\end{split}
\label{eq:energy_sys}
\end{equation}
\begin{equation}
    L_{sys}(\tau)=T_{tot}(\tau),
    \label{eq:latency_sys}
\end{equation}
where, $E_{tot}(\tau), T_{tot}(\tau)$ are derived using Eq.\ref{eq:cost-recursion} (Sec.\ref{sec:sched}), which incorporate the hybrid GainCell-SRAM cache energy (Eq.\ref{eq:gc-energy}) and latency (Eq.\ref{eq:gc-time}) formulations incorporated.


For area derivation, we accumulate the component-wise area of the cache, MMU, NoC routers, DMA, control and SIMD units to obtain the core area $A_{core}$. To build the compute die area $A_{Die}$, we add the area for all its cores along with the D2D router and link areas. The total IO die area $A_{IO}$ comprises of the PHY-modules and D2D interfaces as shown in Fig.\ref{fig:chiplet}(b). We assume that the entire IO die area ($A_{IO}$) is equally distributed across $N_{IO}$ dies which is set equal to twice the number of compute dies along the y-direction as shown in Fig.\ref{fig:chiplet}(b). The system-level area ($A_{sys}$) is derived using the floorplan, described next.

\subsection{Floorplan Generator}
\label{sec:floorplan}
\begin{figure}[h]
    \centering
    \includegraphics[width=\linewidth]{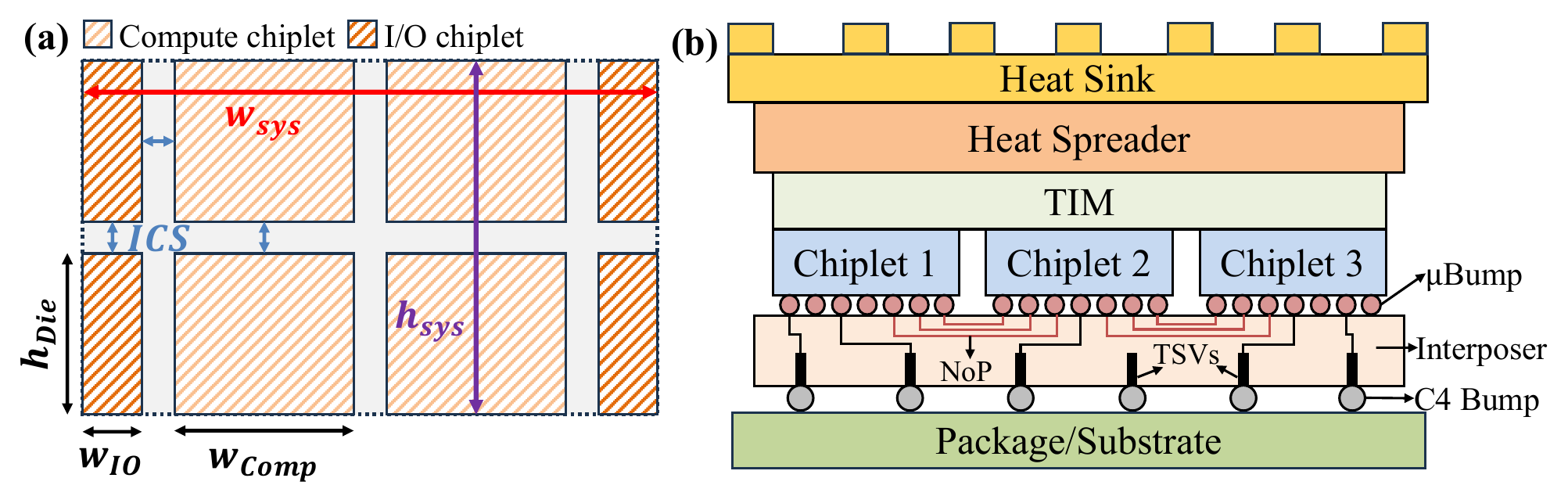}
    \caption{(a) Floorplan and (b) thermal stack of the 2.5D system. }
    \label{fig:floorplan_area}
\end{figure}
Fig.\ref{fig:floorplan_area}(a) shows the floorplan generated by this module. Let us assume that there are $X_c, Y_c$ chiplets along the x- and y-directions respectively. Following \cite{shukla2023temperature}, we assume a uniform inter-chiplet spacing of $ICS$. The compute-die height $h_{Die}$ and width $w_{Comp}$, and the I/O-die width $w_{IO}$, are
\begin{equation}
\resizebox{0.95\columnwidth}{!}{$
    h_{Die}=\sqrt{\frac{y_c}{x_c}A_{Die}}, \quad
    w_{Comp}=\sqrt{\frac{x_c}{y_c}A_{Die}}, \quad
    w_{IO}=\frac{A_{IO}}{2\,h_{Die}\,Y_c}
$}
\end{equation}
From Fig.\ref{fig:floorplan_area}, the system area $A_{sys}$ can be expressed as
\begin{equation}
    A_{sys}=(X_cY_c)A_{Die}+ A_{IO} + (X_c-1)(Y_c-1)ICS
\end{equation}

Above the die layer, heat passes through a single passive stack to ambient, which consists of a thermal interface material (TIM) layer, a heat spreader, and a heat sink (Fig.\ref{fig:floorplan_area}(b)). Each of these is characterized by a thickness, a thermal conductivity (or resistivity), and a volumetric specific heat, as summarized in Table~\ref{tab:thermal-stack}. 

\noindent \textbf{Integration with HotSpot:} For each block in the floorplan, the power is a piecewise trace over the entire execution time ($L_{tot}(\tau)$) than a single averaged value for capturing accurate thermal modeling. During active cycles, a block draws dynamic power given by $P_{block}^{dyn}=E_{block}^{dyn}/t_{active}$,
  where $E_{block}^{dyn}$ is the dynamic energy it accumulates over its active time $t_{active}$. During idle cycles, it draws only its leakage power $P_{block}^{leak}$. This holds uniformly for every
  component in the system. This power-trace along with the above described floorplan is fed to HotSpot (Fig.\ref{fig:overview}) to predict the thermal heatmap for the entire system.

\subsection{Iterative solver}
\label{sec:iter_solver}
Sub-threshold leakage dominates retention-time degradation in eDRAM and gain-cell arrays with temperature \cite{chun20113t}. We model this with an Arrhenius relation,
\begin{equation}
f_t(\theta) = t_{ret}(\theta_{amb})\cdot\exp\!\left(\frac{E_a}{K_B}\left(\frac{1}{\theta}-\frac{1}{\theta_{amb}}\right)\right),
\label{eq:arrhenius}
\end{equation}
giving the retention time $t_{ret}(\theta)=f_t(\theta)$ at die temperature $\theta$, calibrated against the nominal retention time at ambient temperature $\theta_{amb}$, with activation energy $E_a$ and Boltzmann constant $K_B$. The $ELA$ report depends on $t_{ret}$ through $E_X^{refresh}$ (Sec.\ref{sec:refresh}, \ref{sec:ela}), while the floorplan's peak temperature $\theta$ (Sec.~\ref{sec:floorplan}) depends on $ELA$ through the power trace it produces. These two quantities are therefore coupled and must be solved jointly. \begin{algorithm}[h]
  \caption{Iterative ELA--temperature solver}
  \label{alg:iterative-solver}
  \begin{algorithmic}[1]
  \STATE $k \gets 0$, $t_{ret}^{(0)} \gets t_{ret}(\theta_{amb})$, $\theta^{(-1)} \gets \infty$
  \STATE \textbf{repeat}
  \STATE \hspace{1em} compute $ELA^{(k)}$ using $t_{ret}^{(k)}$
  \STATE \hspace{1em} $\theta^{(k)} \gets$ HotSpot (Floorplan Generator, $ELA^{(k)}$)
  \STATE \hspace{1em} $t_{ret}^{(k+1)} \gets f_t\big(\theta^{(k)}\big)$
  \STATE \hspace{1em} $k \gets k+1$
  \STATE \textbf{until} $\big|\theta^{(k-1)}-\theta^{(k-2)}\big| < \epsilon$
  \STATE \textbf{return} $ELA^{(k-1)}$
  \end{algorithmic}
\end{algorithm}
Algorithm \ref{alg:iterative-solver} illustrates the \emph{Iterative solver} algorithm. It alternates computing $ELA$ from the current retention time and updating the retention time from the resulting peak temperature, until the temperature stops changing by more than the tolerance limit $\epsilon$.

\section{Experiments and Results}
\label{sec:exp}
\subsection{Experimental Setup}
\noindent \textbf{2.5D Hardware System:} We use a Simba \cite{shao2019simba}-like 2.5D system at TSMC 40nm for our evaluations, unless stated otherwise. Table \ref{tab:exp_setup_hw} lists the hardware parameters for the 2.5D system, which consists of 36 compute cores divided homogeneously across 9 compute dies. Further, the system is equipped with 6 IO dies. For the hybrid GC-SRAM cache hierarchy, we always incoporate Si-Si GC for $L_1$ and OS-Si GC for $L_2$ cache, as motivated in Sec.\ref{sec:bgnd_motive}.\\

\begin{table}[h]
\centering
\setlength{\fboxsep}{3pt}
\setlength{\fboxrule}{0.6pt}
%
\begin{minipage}[t]{0.48\columnwidth}
\centering
\caption{2.5D accelerator design parameters.}
\label{tab:exp_setup_hw}
\setlength{\tabcolsep}{2pt}
\fbox{%
\resizebox{\dimexpr\linewidth-2\fboxsep-2\fboxrule\relax}{!}{%
\begin{tabular}{@{}ll@{}}
\toprule
\textbf{Parameter} & \textbf{Value} \\
\midrule
$X_c, Y_c$                        & $3, 3$ (9 chiplets) \\
$x_c, y_c$                        & $2, 2$ (4 cores/chiplet) \\
$a, o$                            & $4, 4$ (16 PEs/core) \\
$v$                               & $8$ \\
SIMD throughput                   & 64 OPS/core \\
$S_{AL1}, S_{WL1}, S_{OL1}$       & 8, 4, 2 kB/PE \\
$S_{L2}$                          & 512 kB/core \\
NoC (per-link)                    & 68 GB/s \\
NoP (per-link)                    & 100 GB/s \\
DRAM                              & 72 GB/s (GDDR6X) \\
$\beta_{L_1}^{SRAM}$              & 512 GB/s \\
$\beta_{L_1}^{GC}$                & 128 GB/s \\
$\beta_{L_2}^{SRAM}$              & 256 GB/s \\
$\beta_{L_2}^{GC}$                & 64 GB/s \\
Package                           & Silicon \\
NoP technology                    & UCIe \cite{jain2024design,li2024ucie} \\
Clock frequency                   & 790 MHz \\
\bottomrule
\end{tabular}%
}}
\end{minipage}%
\hfill
\begin{minipage}[t]{0.48\columnwidth}
\centering
\caption{Thermal stack parameters.}
\label{tab:thermal-stack}
\setlength{\tabcolsep}{2pt}
\fbox{%
\resizebox{\dimexpr\linewidth-2\fboxsep-2\fboxrule\relax}{!}{%
\begin{tabular}{@{}ll@{}}
\toprule
\textbf{Parameter} & \textbf{Value} \\
\midrule
die thickness                     & 0.15 mm \\
ICS                               & 2 mm \\
die resistivity                   & $0.01$ m$\cdot$K/W \\
die specific heat                 & $1.75\times10^{6}$ J/m$^3\cdot$K \\
TIM thickness                     & $0.02$ mm \\
TIM conductivity                  & $4.0$ W/m$\cdot$K \\
TIM specific heat                 & $4.0\times10^{6}$ J/m$^3\cdot$K \\
spreader thickness                & $1$ mm \\
sink thickness                    & $6.9$ mm \\
silicon conductivity              & $130$ W/m$\cdot$K \\
copper conductivity               & $400$ W/m$\cdot$K \\
sink convection res.              & $0.4$ K/W \\
sink convection cap.              & $140.4$ J/K \\
ambient temperature               & $313.15$ K \\
\bottomrule
\end{tabular}%
}}
\end{minipage}
\end{table}


\noindent \textbf{Thermal Stack Parameters:} Table \ref{tab:thermal-stack} lists the material properties of the thermal stack comprising of the TIM, heat spreader and heat sink (Fig.\ref{fig:floorplan_area}(b)). For fairness, we use the same parameter values as in \cite{shukla2023temperature}. Also, we use $E_a=0.6(0.8)$ eV (Eq.\ref{eq:arrhenius}) for Si-Si (OS-Si) GC \cite{chun20113t} in Eq.\ref{eq:arrhenius}.\\


\noindent \textbf{Energy and Area Model:} For the MAC array and the MMU (Fig.\ref{fig:core}), we perform energy and area estimations using Synopsys DC \cite{synopsys_dc} at TSMC 40nm with a 790 MHz clock frequency. The communication costs for NoC, NoP and DRAM (including the peripheral IO and PCIe) are obtained using \cite{cai2024gemini}, \cite{wang2025hisim}, \cite{fischer2025floonoc}, and calibrated to TSMC 40nm using \cite{sarangi2021deepscaletool}. Table \ref{tab:energy-constants} shows the cost model for the compute and communication modules.

\begin{table}[h]
\centering
\caption{Energy and area model calibrated to TSMC 40nm.}
\label{tab:energy-constants}
\resizebox{\columnwidth}{!}{%
\begin{tabular}{@{}lrr@{}}
\toprule
\textbf{Component} & \textbf{Energy} & \textbf{Area}\\
\midrule
MAC unit (INT8-INT8)                     & $0.23$ pJ/MAC     & $542.8$ $\mu$m$^2$ \\
NoC \cite{fischer2025floonoc} & $9.264$ pJ/byte/hop     & $28{,}125$ $\mu$m$^2$/(GB/s) \\
NoP \cite{cai2024gemini}  & $2.4$ pJ/byte/hop       & $1056.0$ $\mu$m$^2$/(GB/s) \\
\makecell[l]{DRAM link \\ (GDDR6X) \cite{cai2024gemini}} & $84$ pJ/byte        & \makecell[r]{$116{,}678.6$ $\mu$m$^2$/(GB/s) \\ (PCIe: $23{,}437.5$ $\mu$m$^2$/(GB/s))} \\
\bottomrule
\end{tabular}%
}
\end{table}
Regarding the $L_1$ and $L_2$ caches, the SRAM area and energy are calculated using CACTI \cite{muralimanohar2007cacti6}, while GainCell macros are simulated using OpenGCRAM \cite{wang2025opengcram}. To prevent bank conflicts while feeding to the compute, we set the number of banks in $L_1$ and $L_2$ cache as $v$ and $a\times o$ respectively.\\

\noindent \textbf{LLM Workloads and Models:}
We evaluate five real-world LLM tasks with average prompt and decode token lengths shown in Table \ref{tab:trace_lengths}. We specifically select a mix of prefill- and decode-heavy workloads for comprehensiveness. We evaluate end to end inference for these workloads on the following LLMs: Gemma-2 2B, Llama-3 3B, Qwen2 7B, Llama-3 8B and Mistral-Nemo 12B with W8A8KV8 precision. Batch size is ablated across 8, 32 and 64.

\begin{table}[h]
\centering
\caption{Average token lengths of LLM workloads used in this work.}
\label{tab:trace_lengths}
\begin{tabular}{l|c|c}
\hline
\textbf{Workload} &
\textbf{Prompt length} &
\textbf{Decode length} \\
\hline

Azure Code \cite{azurepublicdataset} &
2048 &
28 \\

openchat\_sharegpt4 \cite{agrawal2024sarathi} &
1730 &
415 \\

arxiv\_summarization \cite{agrawal2024sarathi} &
7059 &
208 \\

OpenCodeInstruct \cite{ahmad2025opencodeinstruct} &
1340 &
2150 \\

ShareChat \cite{yan2025sharechat} &
519 &
1115 \\
\hline

\end{tabular}%
\end{table} 

\subsection{Data Lifetime Analysis}
\begin{figure*}[h]
    \centering
    \includegraphics[width=\linewidth]{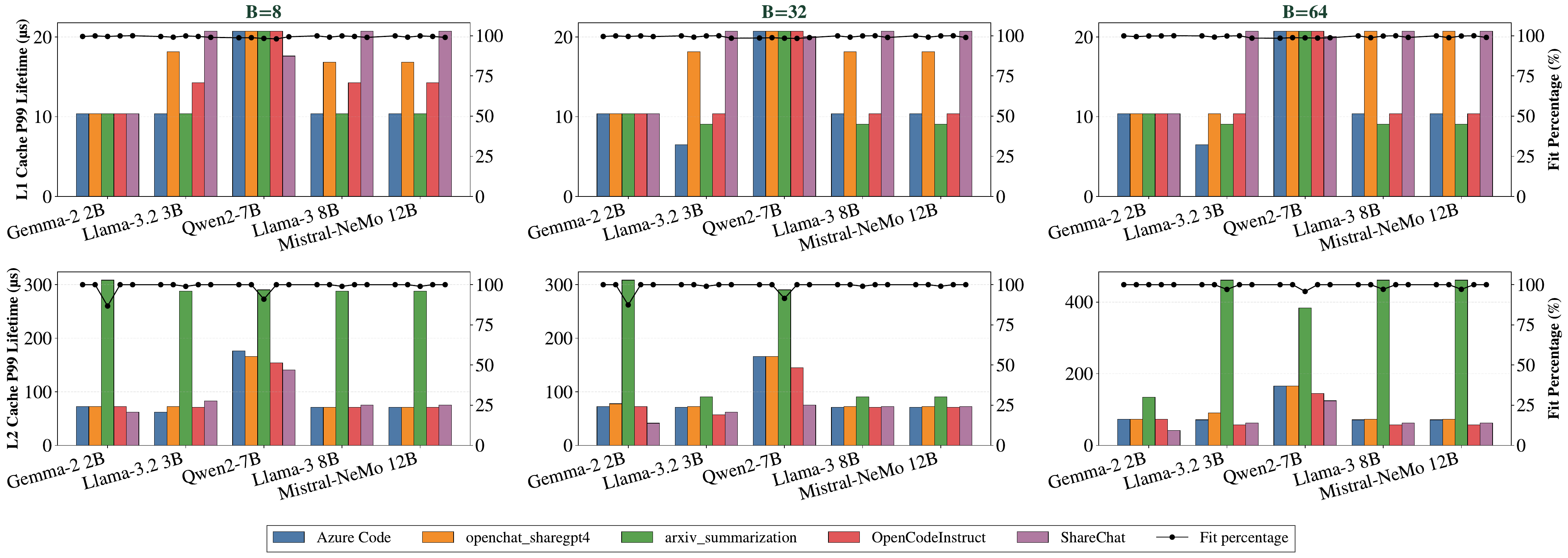}
    \caption{Data lifetime statistics for prefill stage across different LLM models and workloads.}
    \label{fig:data_prefill}
\end{figure*}

\begin{figure*}[h]
    \centering
    \includegraphics[width=\linewidth]{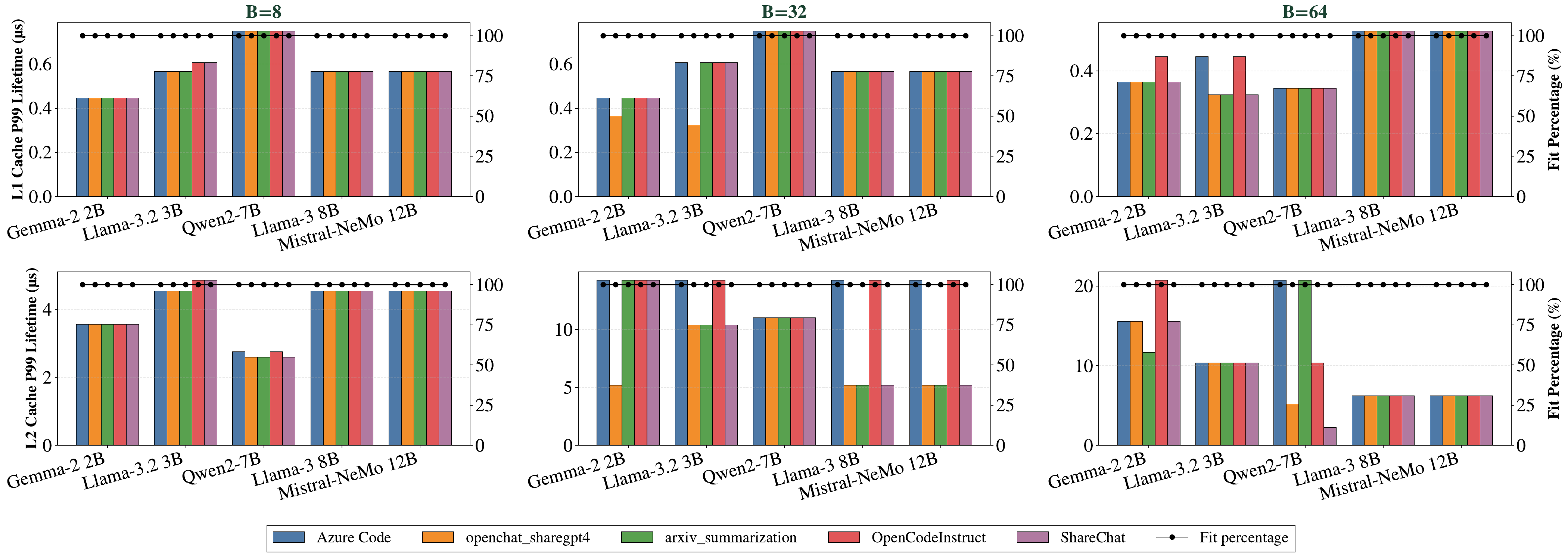}
    \caption{Data lifetime statistics for decode stage across different LLM models and workloads.}
    \label{fig:data_decode}
\end{figure*}

Fig.\ref{fig:data_prefill} and \ref{fig:data_decode} show the prefill and decode stage data lifetime distributions. We define a metric called \textit{Fit percentage}, which denotes the fraction of data tiles with lifetime lower than 20$\mu$s and 200$\mu$s for $L_1$ and $L_2$ cache respectively. 
\begin{equation}
\resizebox{\columnwidth}{!}{$
\underset{X\in\{L_1,L_2\}}{\mathrm{Fit\%}_X}=\frac{\#\,\text{tiles with } \lambda<t_{ret,X}}{\#\,\text{tiles}}\times 100,\quad t_{ret,L_1}=20\,\mu\text{s},\ t_{ret,L_2}=200\,\mu\text{s}
$}
\end{equation}
Higher the fit percentage, lower will be the refresh energy overhead as more tiles will fit within the GainCell caches. Clearly, prefill stage has higher P99 data lifetime compared to decode stage as it involves more number of computations. However, the fit percentage is still close to 100\% for the prefill stage across most of the workloads at both the caches, indicating that GainCells can be used to replace SRAM to a large extent. Decode stage is found to be not a concern regarding limited data retention as it always achieves 100\% fit percentage.

\begin{figure}[h]
    \centering
    \includegraphics[width=\linewidth]{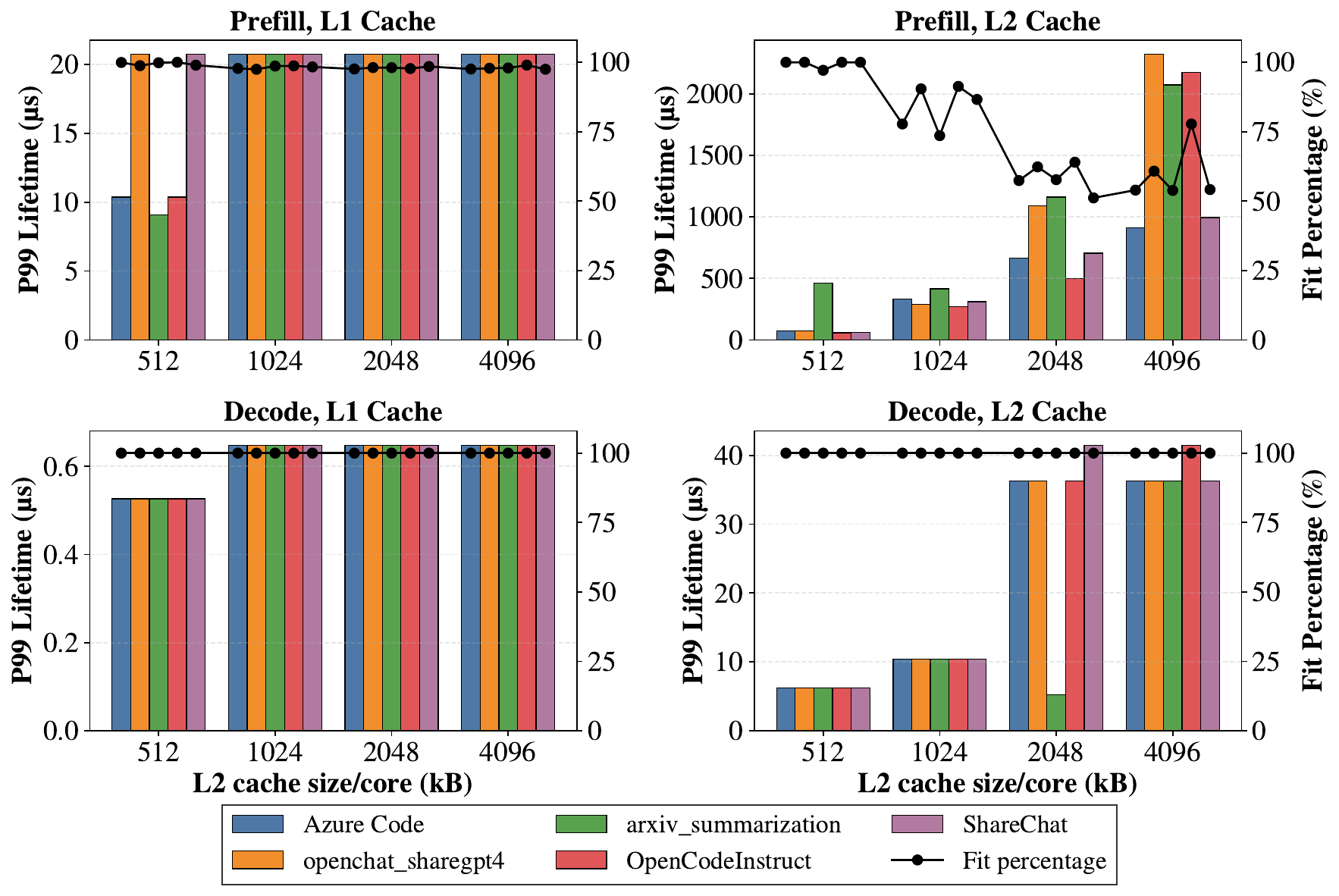}
    \caption{Ablation of data lifetime for LLama-3 8B at batch size 64 for different L2 cache sizes.}
    \label{fig:data_lifetime_ablate}
\end{figure}

Since batch size ($B$) 64 achieves the highest data lifetimes, we perform a worst-case lifetime analysis for $B=64$. Fig.\ref{fig:data_lifetime_ablate} ablates the impact of increasing $L_2$ cache size on data lifetimes. It can be easily concluded that only the $L_2$ cache data lifetime during the prefill stage increases significantly with larger cache and drastically drops the fit percentage to even below 50\% in some cases. This is expected since larger cache allows better data reuse by staging the data longer. Hence, while at smaller $L_2$ cache capacities, we can replace SRAM completely with 2T OS-Si GC, doing so at larger cache sizes could incur considerable refresh energy, as shown later in Sec.\ref{sec:hw_eff_result}.

\subsection{ELA Evaluation}

\subsubsection{Thermal simulation} 
\begin{figure}[h]
    \centering
    \includegraphics[width=\linewidth]{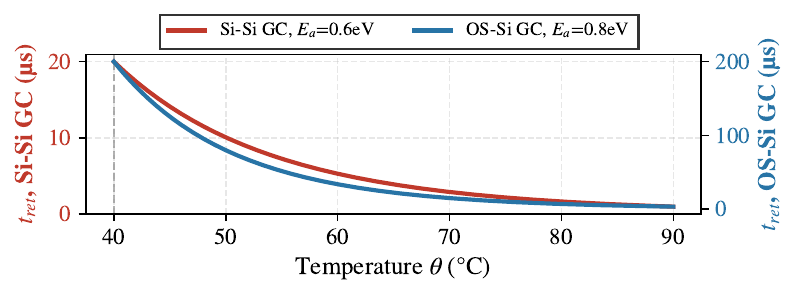}
    \caption{Variation of retention periods of Si-Si and OS-Si GC with temperature.}
    \label{fig:arrhenius}
\end{figure}

\begin{figure}[h]
    \centering
    \includegraphics[width=\linewidth]{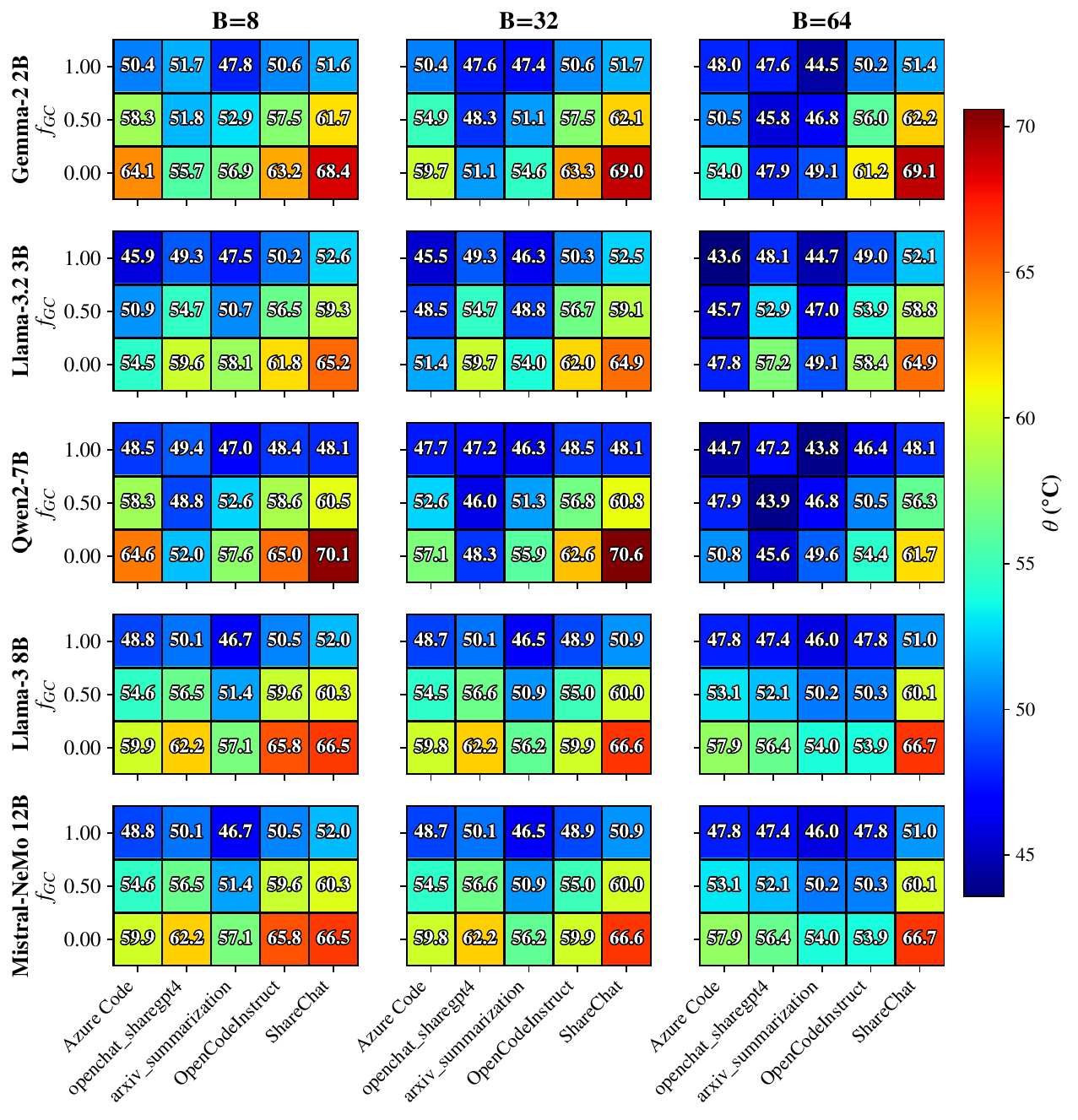}
    \caption{Peak temperature for different workloads during the prefill stage.}
    \label{fig:thermal_prefill}
\end{figure}

\begin{figure}[h]
    \centering
    \includegraphics[width=\linewidth]{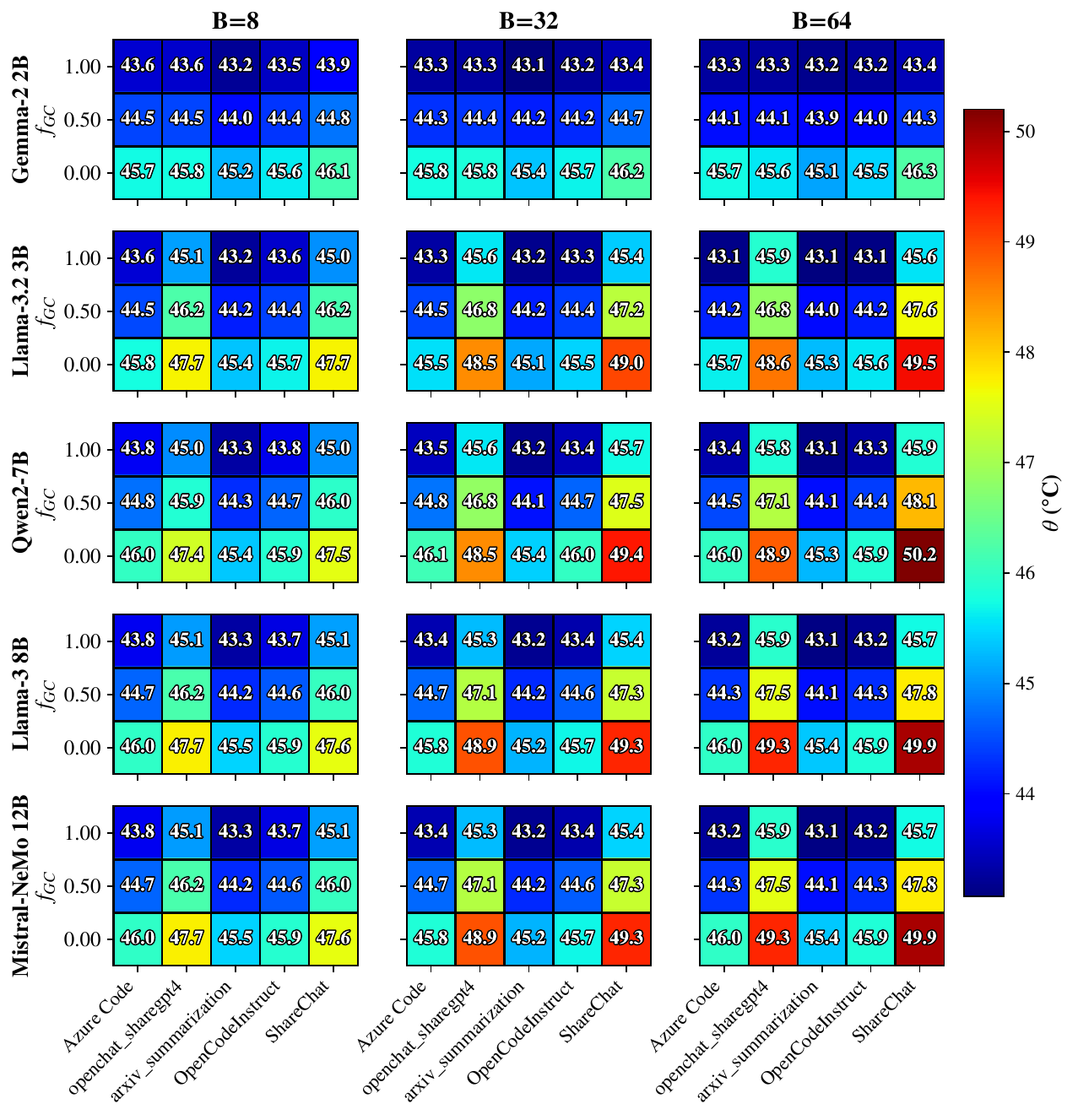}
    \caption{Peak temperature for different workloads during the decode stage.}
    \label{fig:thermal_decode}
\end{figure}

Fig.\ref{fig:arrhenius} shows how the data retention period varies with temperature for the 2T GCs, based on Eq.\ref{eq:arrhenius}.
Next, we vary the GC fraction ($f_{GC,L_1}=f_{GC,L_2}=f_{GC}$) from 0 (pure SRAM cache) to 1 (pure GainCell cache) for calculating the system-level ELA metrics with GainCell caches. This is needed to report the end-to-end inference energy and latency faithfully as they depend on the GC retention period, which in turn depends on the temperature (Sec.\ref{sec:iter_solver}). Fig.\ref{fig:thermal_prefill} and \ref{fig:thermal_decode} show the steady-state peak temperature during the prefill and decode stages respectively across different models and workloads, obtained using Iterative solver (Sec.\ref{sec:iter_solver}). Since prefill involves more computation and cache accesses than decode, it results in higher die temperature. However, the peak temperature monotonically reduces as $f_{GC}$ is increased, indicating that higher $f_{GC}$ is also associated with a lower system power. This follows from Fig.\ref{fig:data_prefill} and \ref{fig:data_decode} as the refresh energy would be negligible due to such high fit percentages (Eq.\ref{eq:refresh-energy}), resulting in lower power due to energy-efficiency of the GCs alone. Let2Gain finds that GainCell caches ($f_{GC}=1$) lower the peak temperature of the system by up to $\mathbf{17^{\circ}C}$ and $\mathbf{4.2^{\circ}C}$ during the prefill and decode stages respectively.

\subsubsection{Hardware efficiency and throughput trade-off}
\label{sec:hw_eff_result}
\begin{figure*}[h]
    \centering
    \includegraphics[width=\linewidth]{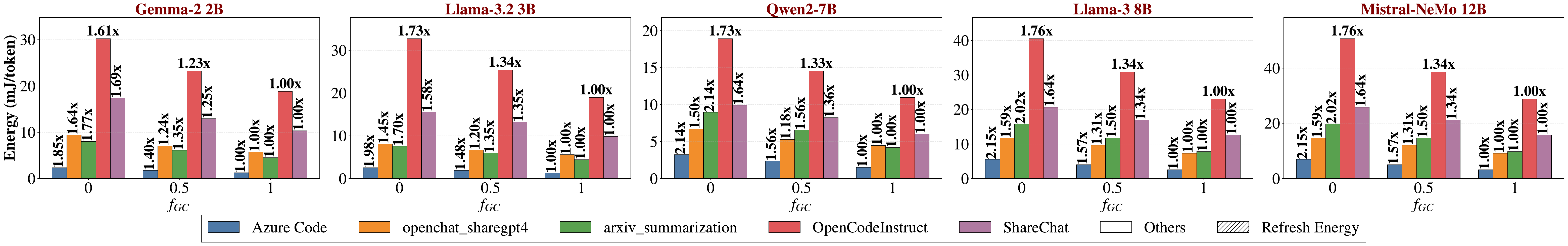}
    \caption{Inference energy reduction with increase in GainCell fraction ($f_{GC}$) across different models and workloads for batch size, $B=64$. Refresh energy overhead is $<1\%$ due to most of the data tiles fitting within the retention limits, as shown in Fig.\ref{fig:data_prefill} and \ref{fig:data_decode} from P99 lifetimes and fit percentages.}
    \label{fig:energy_infer}
\end{figure*}

Fig.\ref{fig:energy_infer} shows the end to end inference energy improvements across different workloads as $f_{GC}$ is increased on the hardware defined in Table \ref{tab:exp_setup_hw}. Let2Gain finds that by replacing the entire SRAM stack with GainCells ($f_{GC}=1$), inference energy improves by up to $\mathbf{2.15\times}$.

Next, we study the impact of slower GainCell cache on the system throughput. We define an efficiency metric as \texttt{tokens per second/(Watt-mm$^2$}) (higher is better), which is throughput normalized to power times area. Table \ref{tab:gc_frac_efficiency} shows how the efficiency, TTFT and total runtime vary with $f_{GC}$ across different workloads. Maximum efficiency is achieved always at $f_{GC}=1$ (pure GainCell cache) for every workload. Let2Gain identifies up to $\mathbf{2.83\times}$ higher end-to-end efficiency over a hardware with SRAM-based cache. 

\begin{figure}[h]
    \centering
    \includegraphics[width=\linewidth]{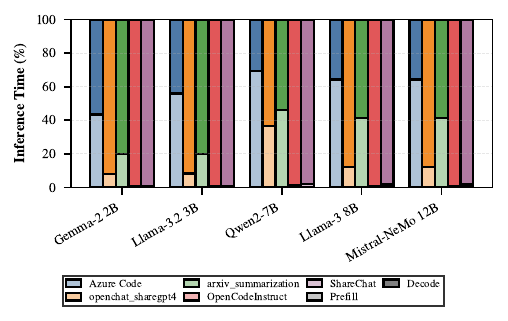}
    \caption{End to end inference runtime split between prefill and decode stages at batch size 32 on Table \ref{tab:exp_setup_hw} hardware.}
    \label{fig:time_split}
\end{figure}

However, the slower GainCell caches often result in increased TTFT (marked in \textcolor{red}{red} in Table \ref{tab:gc_frac_efficiency}), up to $\mathbf{1.88\times}$ over SRAM baseline. This is a fundamental limitation of GainCells over SRAM, and affects the latency when  $T_X(\sigma_\ell)$ in Eq.\ref{eq:time} becomes the bottleneck due to the slower read bandwidth of GainCells (Table \ref{tab:memory_comparison}). However, total runtime rises by at most $\mathbf{21\%}$, as decode dominates most workloads (Fig.~\ref{fig:time_split}). In some cases, $f_{GC}=1$ even lowers TTFT (\textcolor{blue}{blue}), since the Mapper and Scheduler (Sec.~\ref{sec:sched}) accounts for the GainCell fraction and finds schedules different from pure SRAM.
\begin{table*}[h]
\centering
\caption{Impact of $f_{GC}$ on end-to-end efficiency (tokens/s/(W$\cdot$mm$^2$)) ($\uparrow$ is better), TTFT, and total runtime ($\downarrow$ is better) normalized to $f_{GC}=0$ (pure SRAM cache) per (workload, model, batch size). Time between tokens (TBT) is omitted as it showed no variation with $f_{GC}$ across all configurations.}
\label{tab:gc_frac_efficiency}
\resizebox{\textwidth}{!}{%
\begin{tabular}{llcccccccccccccccccc}
\toprule
\multirow{3}{*}{\textbf{Workload}} & \multirow{3}{*}{\textbf{Model}} & \multicolumn{6}{c}{\textbf{Efficiency}} & \multicolumn{6}{c}{\textbf{TTFT}} & \multicolumn{6}{c}{\textbf{Total Runtime}} \\
\cmidrule(lr){3-8} \cmidrule(lr){9-14} \cmidrule(lr){15-20}
 & & \multicolumn{3}{c}{$f_{GC}=0.5$} & \multicolumn{3}{c}{$f_{GC}=1$} & \multicolumn{3}{c}{$f_{GC}=0.5$} & \multicolumn{3}{c}{$f_{GC}=1$} & \multicolumn{3}{c}{$f_{GC}=0.5$} & \multicolumn{3}{c}{$f_{GC}=1$} \\
\cmidrule(lr){3-5} \cmidrule(lr){6-8} \cmidrule(lr){9-11} \cmidrule(lr){12-14} \cmidrule(lr){15-17} \cmidrule(lr){18-20}
 & & $B{=}8$ & $B{=}32$ & $B{=}64$ & $B{=}8$ & $B{=}32$ & $B{=}64$ & $B{=}8$ & $B{=}32$ & $B{=}64$ & $B{=}8$ & $B{=}32$ & $B{=}64$ & $B{=}8$ & $B{=}32$ & $B{=}64$ & $B{=}8$ & $B{=}32$ & $B{=}64$ \\
\midrule
\multirow{5}{*}{Azure Code} & Gemma-2 2B & \textcolor{blue}{1.45} & \textcolor{blue}{1.48} & \textcolor{blue}{1.50} & \textcolor{blue}{2.28} & \textcolor{blue}{2.39} & \textcolor{blue}{2.44} & 1.00 & 1.00 & 1.00 & \textcolor{red}{1.20} & \textcolor{red}{1.05} & \textcolor{blue}{0.87} & 1.00 & 1.00 & 1.00 & \textcolor{red}{1.07} & \textcolor{red}{1.02} & \textcolor{blue}{0.93} \\
 & Llama-3.2 3B & \textcolor{blue}{1.46} & \textcolor{blue}{1.47} & \textcolor{blue}{1.52} & \textcolor{blue}{2.32} & \textcolor{blue}{2.44} & \textcolor{blue}{2.61} & 1.00 & 1.00 & 1.00 & \textcolor{red}{1.24} & \textcolor{red}{1.02} & 1.00 & 1.00 & 1.00 & 1.00 & \textcolor{red}{1.11} & \textcolor{red}{1.01} & 1.00 \\
 & Qwen2-7B & \textcolor{blue}{1.51} & \textcolor{blue}{1.53} & \textcolor{blue}{1.55} & \textcolor{blue}{2.56} & \textcolor{blue}{2.71} & \textcolor{blue}{2.81} & 1.00 & 1.00 & 1.00 & \textcolor{red}{1.37} & \textcolor{red}{1.07} & 1.00 & 1.00 & 1.00 & 1.00 & \textcolor{red}{1.21} & \textcolor{red}{1.05} & 1.00 \\
 & Llama-3 8B & \textcolor{blue}{1.53} & \textcolor{blue}{1.54} & \textcolor{blue}{1.56} & \textcolor{blue}{2.69} & \textcolor{blue}{2.77} & \textcolor{blue}{2.83} & 1.00 & 1.00 & 1.00 & 1.00 & 1.00 & 1.00 & 1.00 & 1.00 & 1.00 & 1.00 & 1.00 & 1.00 \\
 & Mistral-NeMo 12B & \textcolor{blue}{1.53} & \textcolor{blue}{1.54} & \textcolor{blue}{1.56} & \textcolor{blue}{2.69} & \textcolor{blue}{2.77} & \textcolor{blue}{2.83} & 1.00 & 1.00 & 1.00 & 1.00 & 1.00 & 1.00 & 1.00 & 1.00 & 1.00 & 1.00 & 1.00 & 1.00 \\
\midrule
\multirow{5}{*}{openchat\_sharegpt4} & Gemma-2 2B & \textcolor{blue}{1.39} & \textcolor{blue}{1.45} & \textcolor{blue}{1.50} & \textcolor{blue}{2.05} & \textcolor{blue}{2.17} & \textcolor{blue}{2.16} & 1.00 & 1.00 & 1.00 & \textcolor{blue}{0.77} & \textcolor{blue}{0.97} & \textcolor{blue}{0.62} & 1.00 & 1.00 & 1.00 & \textcolor{blue}{0.99} & 1.00 & \textcolor{blue}{0.95} \\
 & Llama-3.2 3B & \textcolor{blue}{1.34} & \textcolor{blue}{1.37} & \textcolor{blue}{1.38} & \textcolor{blue}{1.95} & \textcolor{blue}{1.95} & \textcolor{blue}{1.91} & 1.00 & 1.00 & 1.00 & 1.00 & 1.00 & 1.00 & 1.00 & 1.00 & 1.00 & 1.00 & 1.00 & 1.00 \\
 & Qwen2-7B & \textcolor{blue}{1.39} & \textcolor{blue}{1.41} & \textcolor{blue}{1.44} & \textcolor{blue}{1.93} & \textcolor{blue}{1.81} & \textcolor{blue}{1.98} & 1.00 & 1.00 & 1.00 & \textcolor{blue}{0.77} & \textcolor{blue}{0.96} & \textcolor{blue}{0.51} & 1.00 & 1.00 & 1.00 & \textcolor{blue}{0.96} & \textcolor{blue}{0.99} & \textcolor{blue}{0.73} \\
 & Llama-3 8B & \textcolor{blue}{1.36} & \textcolor{blue}{1.39} & \textcolor{blue}{1.38} & \textcolor{blue}{2.02} & \textcolor{blue}{2.22} & \textcolor{blue}{2.09} & 1.00 & 1.00 & 1.00 & \textcolor{red}{1.02} & \textcolor{red}{1.02} & 1.00 & 1.00 & 1.00 & 1.00 & 1.00 & 1.00 & 1.00 \\
 & Mistral-NeMo 12B & \textcolor{blue}{1.36} & \textcolor{blue}{1.39} & \textcolor{blue}{1.38} & \textcolor{blue}{2.02} & \textcolor{blue}{2.22} & \textcolor{blue}{2.09} & 1.00 & 1.00 & 1.00 & \textcolor{red}{1.02} & \textcolor{red}{1.02} & 1.00 & 1.00 & 1.00 & 1.00 & 1.00 & 1.00 & 1.00 \\
\midrule
\multirow{5}{*}{arxiv\_summarization} & Gemma-2 2B & \textcolor{blue}{1.43} & \textcolor{blue}{1.42} & \textcolor{blue}{1.49} & \textcolor{blue}{2.17} & \textcolor{blue}{2.27} & \textcolor{blue}{2.33} & 1.00 & 1.00 & 1.00 & \textcolor{red}{1.13} & \textcolor{red}{1.07} & 1.00 & 1.00 & 1.00 & 1.00 & \textcolor{red}{1.02} & \textcolor{red}{1.01} & 1.00 \\
 & Llama-3.2 3B & \textcolor{blue}{1.40} & \textcolor{blue}{1.40} & \textcolor{blue}{1.44} & \textcolor{blue}{2.02} & \textcolor{blue}{2.04} & \textcolor{blue}{2.24} & \textcolor{red}{1.34} & \textcolor{red}{1.28} & 1.00 & \textcolor{red}{1.42} & \textcolor{red}{1.35} & 1.00 & \textcolor{red}{1.05} & \textcolor{red}{1.06} & 1.00 & \textcolor{red}{1.07} & \textcolor{red}{1.07} & 1.00 \\
 & Qwen2-7B & \textcolor{blue}{1.53} & \textcolor{blue}{1.54} & \textcolor{blue}{1.56} & \textcolor{blue}{2.67} & \textcolor{blue}{2.72} & \textcolor{blue}{2.82} & 1.00 & 1.00 & 1.00 & \textcolor{red}{1.07} & \textcolor{red}{1.07} & 1.00 & 1.00 & 1.00 & 1.00 & \textcolor{red}{1.03} & \textcolor{red}{1.03} & 1.00 \\
 & Llama-3 8B & \textcolor{blue}{1.50} & \textcolor{blue}{1.51} & \textcolor{blue}{1.53} & \textcolor{blue}{2.49} & \textcolor{blue}{2.54} & \textcolor{blue}{2.66} & \textcolor{red}{1.10} & \textcolor{red}{1.10} & 1.00 & \textcolor{red}{1.12} & \textcolor{red}{1.12} & 1.00 & \textcolor{red}{1.04} & \textcolor{red}{1.04} & 1.00 & \textcolor{red}{1.05} & \textcolor{red}{1.05} & 1.00 \\
 & Mistral-NeMo 12B & \textcolor{blue}{1.50} & \textcolor{blue}{1.51} & \textcolor{blue}{1.53} & \textcolor{blue}{2.49} & \textcolor{blue}{2.54} & \textcolor{blue}{2.66} & \textcolor{red}{1.10} & \textcolor{red}{1.10} & 1.00 & \textcolor{red}{1.12} & \textcolor{red}{1.12} & 1.00 & \textcolor{red}{1.04} & \textcolor{red}{1.04} & 1.00 & \textcolor{red}{1.05} & \textcolor{red}{1.05} & 1.00 \\
\midrule
\multirow{5}{*}{OpenCodeInstruct} & Gemma-2 2B & \textcolor{blue}{1.38} & \textcolor{blue}{1.44} & \textcolor{blue}{1.48} & \textcolor{blue}{2.03} & \textcolor{blue}{2.19} & \textcolor{blue}{2.11} & 1.00 & 1.00 & 1.00 & \textcolor{red}{1.06} & \textcolor{red}{1.06} & \textcolor{red}{1.06} & 1.00 & 1.00 & 1.00 & 1.00 & 1.00 & 1.00 \\
 & Llama-3.2 3B & \textcolor{blue}{1.38} & \textcolor{blue}{1.38} & \textcolor{blue}{1.46} & \textcolor{blue}{2.01} & \textcolor{blue}{2.08} & \textcolor{blue}{2.27} & 1.00 & 1.00 & 1.00 & \textcolor{red}{1.07} & \textcolor{red}{1.07} & 1.00 & 1.00 & 1.00 & 1.00 & 1.00 & 1.00 & 1.00 \\
 & Qwen2-7B & \textcolor{blue}{1.37} & \textcolor{blue}{1.39} & \textcolor{blue}{1.48} & \textcolor{blue}{1.99} & \textcolor{blue}{2.17} & \textcolor{blue}{2.27} & 1.00 & 1.00 & 1.00 & \textcolor{red}{1.39} & \textcolor{red}{1.40} & 1.00 & 1.00 & 1.00 & 1.00 & 1.00 & 1.00 & 1.00 \\
 & Llama-3 8B & \textcolor{blue}{1.38} & \textcolor{blue}{1.38} & \textcolor{blue}{1.49} & \textcolor{blue}{2.00} & \textcolor{blue}{2.12} & \textcolor{blue}{2.31} & 1.00 & 1.00 & 1.00 & \textcolor{red}{1.29} & \textcolor{red}{1.23} & \textcolor{blue}{0.93} & 1.00 & 1.00 & 1.00 & 1.00 & 1.00 & 1.00 \\
 & Mistral-NeMo 12B & \textcolor{blue}{1.38} & \textcolor{blue}{1.38} & \textcolor{blue}{1.49} & \textcolor{blue}{2.00} & \textcolor{blue}{2.12} & \textcolor{blue}{2.31} & 1.00 & 1.00 & 1.00 & \textcolor{red}{1.29} & \textcolor{red}{1.23} & \textcolor{blue}{0.93} & 1.00 & 1.00 & 1.00 & 1.00 & 1.00 & 1.00 \\
\midrule
\multirow{5}{*}{ShareChat} & Gemma-2 2B & \textcolor{blue}{1.37} & \textcolor{blue}{1.43} & \textcolor{blue}{1.53} & \textcolor{blue}{1.99} & \textcolor{blue}{2.28} & \textcolor{blue}{2.22} & 1.00 & 1.00 & 1.00 & \textcolor{red}{1.29} & \textcolor{red}{1.30} & \textcolor{red}{1.30} & 1.00 & 1.00 & 1.00 & 1.00 & 1.00 & 1.00 \\
 & Llama-3.2 3B & \textcolor{blue}{1.33} & \textcolor{blue}{1.35} & \textcolor{blue}{1.34} & \textcolor{blue}{1.94} & \textcolor{blue}{2.05} & \textcolor{blue}{2.08} & 1.00 & 1.00 & 1.00 & \textcolor{red}{1.05} & \textcolor{red}{1.05} & 1.00 & 1.00 & 1.00 & 1.00 & 1.00 & 1.00 & 1.00 \\
 & Qwen2-7B & \textcolor{blue}{1.35} & \textcolor{blue}{1.36} & \textcolor{blue}{1.37} & \textcolor{blue}{1.92} & \textcolor{blue}{2.04} & \textcolor{blue}{2.16} & \textcolor{red}{1.12} & \textcolor{red}{1.12} & 1.00 & \textcolor{red}{1.86} & \textcolor{red}{1.88} & \textcolor{red}{1.50} & 1.00 & 1.00 & 1.00 & \textcolor{red}{1.01} & \textcolor{red}{1.02} & \textcolor{red}{1.02} \\
 & Llama-3 8B & \textcolor{blue}{1.34} & \textcolor{blue}{1.38} & \textcolor{blue}{1.39} & \textcolor{blue}{1.87} & \textcolor{blue}{2.11} & \textcolor{blue}{2.15} & 1.00 & 1.00 & 1.00 & \textcolor{red}{1.19} & \textcolor{red}{1.17} & \textcolor{red}{1.17} & 1.00 & 1.00 & 1.00 & 1.00 & 1.00 & \textcolor{red}{1.01} \\
 & Mistral-NeMo 12B & \textcolor{blue}{1.34} & \textcolor{blue}{1.38} & \textcolor{blue}{1.39} & \textcolor{blue}{1.87} & \textcolor{blue}{2.11} & \textcolor{blue}{2.15} & 1.00 & 1.00 & 1.00 & \textcolor{red}{1.19} & \textcolor{red}{1.17} & \textcolor{red}{1.17} & 1.00 & 1.00 & 1.00 & 1.00 & 1.00 & \textcolor{red}{1.01} \\
\bottomrule
\end{tabular}%
}
\end{table*}
\begin{figure}[h]
    \centering
    \includegraphics[width=\linewidth]{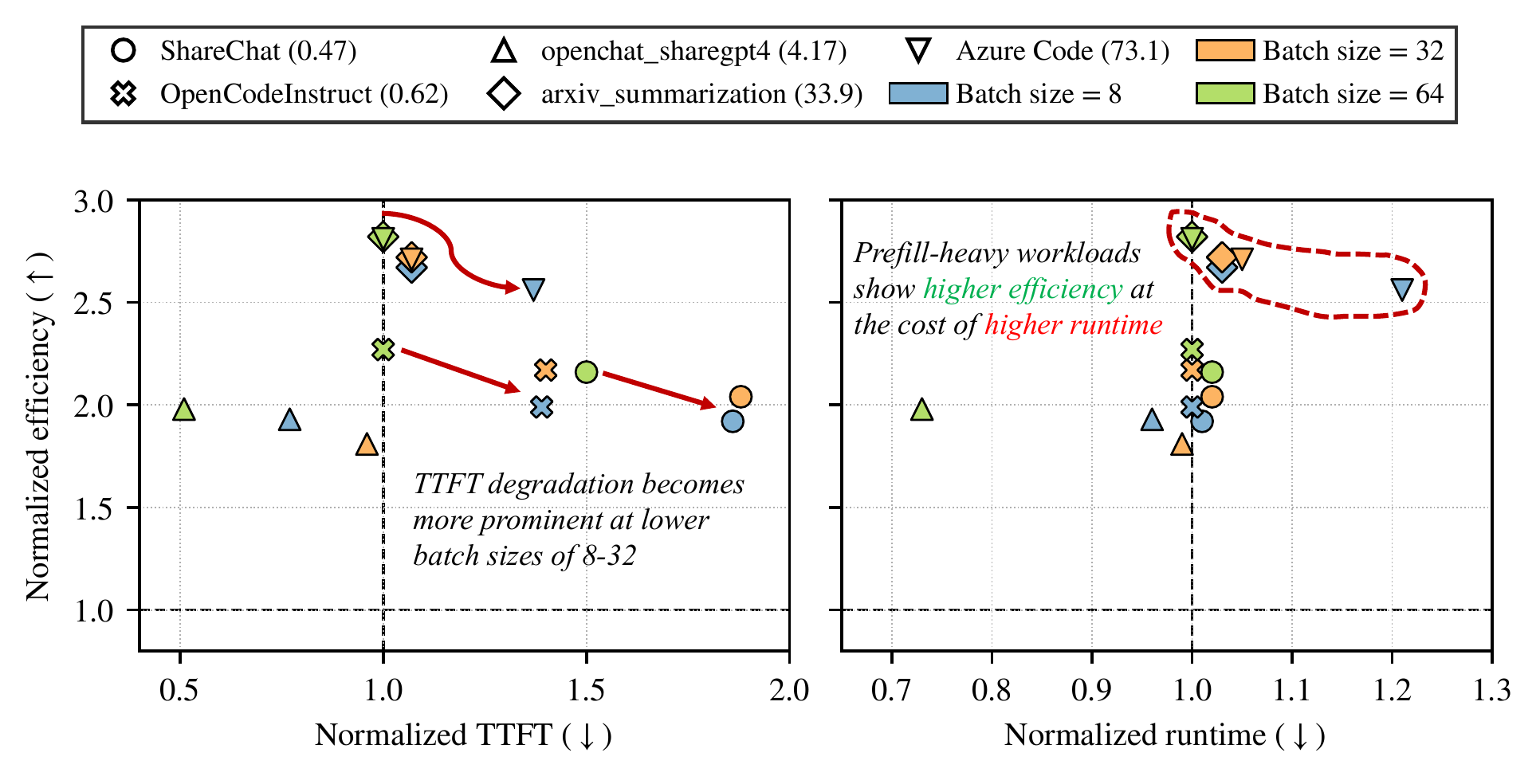}
    \caption{Efficiency–latency trade-off of Qwen2-7B across batch sizes and workloads (prefill-to-decode token ratio in parentheses) for $f_{GC}=1$, normalized to pure SRAM.}
    \label{fig:scaling_trend}
\end{figure}
Fig.\ref{fig:scaling_trend} concludes that the prefill-heavy workloads achieve higher efficiency with GainCells albeit at the cost of higher TTFT/runtime degradation. Runtime degradation is larger at low batch sizes, where low-bandwidth GC cache access is the bottleneck as most of the data fits on-chip. Higher batch sizes show off-chip spillage, making DRAM access the bottleneck.

\noindent \textbf{Ablation study on $L_2$ cache size:} Since Fig.\ref{fig:data_lifetime_ablate} showed that the fit percentage decayed with larger $L_2$ caches, we aim to evaluate the refresh energy overhead variation and its impact on end-to-end performance with $L_2$ cache size. At batch size 32, we compare $f_{GC}=1$ system with $f_{GC}=0$ system at different per core $L_2$ cache size. Table \ref{tab:ubuf_ablation} summarizes our findings. Among the evaluated workloads, the prefill-heavy tasks (\texttt{Azure Code}, \texttt{openchat\_sharegpt4}, and \texttt{arxiv\_summarization}) exhibit the highest refresh energy overhead, reaching $\mathbf{38.8\%}$ at a 4~MB $L_2$ cache size, whereas the decode-heavy tasks show negligible overhead. Further, the runtime penalty with GCs becomes more pronounced at larger cache sizes, with slowdown reaching up to $\mathbf{1.92\times}$. Nevertheless, larger cache configurations benefit the most from GainCell integration, achieving an efficiency gain of up to $\mathbf{5.03\times}$ over SRAM, a result of SRAM's leakage energy scaling unfavorably at these sizes. These results indicate that although refresh overhead is non-trivial for prefill-heavy workloads, the associated energy and area savings more than compensate for it. In conclusion, GainCell can replace SRAM-based cache hierarchy fully while providing an efficiency-throughput trade-off. \textit{We therefore perform design space exploration (DSE) to find the best hybrid GC-SRAM cache composition for optimizing different objectives, discussed next.}

\begin{table*}[h]
\centering
\caption{L2 cache size (per core) ablation ($B$=32). Refresh energy (\% of total energy), total runtime and efficiency (tokens/s/(W$\cdot$mm$^2$)) at $f_{GC}=1$, normalized to $f_{GC}=0$ (pure SRAM cache) per (workload, model, L2 size).}
\label{tab:ubuf_ablation}
\resizebox{\textwidth}{!}{%
\begin{tabular}{llccccccccc}
\toprule
\multirow{2}{*}{\textbf{Workload}} & \multirow{2}{*}{\textbf{Model}} & \multicolumn{3}{c}{\textbf{Refresh Energy}} & \multicolumn{3}{c}{\textbf{Total Runtime}} & \multicolumn{3}{c}{\textbf{Efficiency}} \\
\cmidrule(lr){3-5} \cmidrule(lr){6-8} \cmidrule(lr){9-11}
 & & $L_2=1$MB & $L_2=2$MB & $L_2=4$MB & $L_2=1$MB & $L_2=2$MB & $L_2=4$MB & $L_2=1$MB & $L_2=2$MB & $L_2=4$MB \\
\midrule
\multirow{5}{*}{Azure Code} & Gemma-2 2B & 4.2\% & 25.2\% & 1.4\% & \textcolor{red}{1.08} & \textcolor{red}{1.12} & \textcolor{red}{1.11} & \textcolor{blue}{2.69} & \textcolor{blue}{2.25} & \textcolor{blue}{4.09} \\
 & Llama-3.2 3B & 0.5\% & 20.9\% & 22.8\% & \textcolor{red}{1.07} & \textcolor{red}{1.04} & \textcolor{red}{1.05} & \textcolor{blue}{2.66} & \textcolor{blue}{2.43} & \textcolor{blue}{3.22} \\
 & Qwen2-7B & 2.9\% & 1.6\% & 0.3\% & \textcolor{red}{1.31} & \textcolor{red}{1.66} & \textcolor{red}{1.92} & \textcolor{blue}{2.72} & \textcolor{blue}{3.15} & \textcolor{blue}{3.47} \\
 & Llama-3 8B & 3.6\% & 7.4\% & 28.9\% & \textcolor{red}{1.15} & \textcolor{red}{1.40} & \textcolor{red}{1.29} & \textcolor{blue}{2.76} & \textcolor{blue}{2.88} & \textcolor{blue}{2.83} \\
 & Mistral-NeMo 12B & 3.6\% & 7.4\% & 28.9\% & \textcolor{red}{1.15} & \textcolor{red}{1.40} & \textcolor{red}{1.29} & \textcolor{blue}{2.76} & \textcolor{blue}{2.88} & \textcolor{blue}{2.83} \\
\midrule
\multirow{5}{*}{openchat\_sharegpt4} & Gemma-2 2B & 0.4\% & 1.0\% & 5.2\% & \textcolor{red}{1.01} & \textcolor{red}{1.01} & \textcolor{red}{1.02} & \textcolor{blue}{2.37} & \textcolor{blue}{3.71} & \textcolor{blue}{3.84} \\
 & Llama-3.2 3B & 0.0\% & 2.7\% & 16.9\% & \textcolor{red}{1.01} & 1.00 & \textcolor{red}{1.01} & \textcolor{blue}{2.30} & \textcolor{blue}{3.52} & \textcolor{blue}{2.89} \\
 & Qwen2-7B & 0.5\% & 3.7\% & 8.2\% & 1.00 & \textcolor{red}{1.13} & \textcolor{red}{1.13} & \textcolor{blue}{2.90} & \textcolor{blue}{2.93} & \textcolor{blue}{3.52} \\
 & Llama-3 8B & 0.5\% & 2.5\% & 7.7\% & \textcolor{red}{1.04} & \textcolor{red}{1.03} & \textcolor{red}{1.05} & \textcolor{blue}{2.42} & \textcolor{blue}{3.51} & \textcolor{blue}{3.81} \\
 & Mistral-NeMo 12B & 0.5\% & 2.5\% & 7.7\% & \textcolor{red}{1.04} & \textcolor{red}{1.03} & \textcolor{red}{1.05} & \textcolor{blue}{2.42} & \textcolor{blue}{3.51} & \textcolor{blue}{3.81} \\
\midrule
\multirow{5}{*}{arxiv\_summarization} & Gemma-2 2B & 1.2\% & 3.4\% & 36.7\% & 1.00 & \textcolor{red}{1.02} & \textcolor{red}{1.03} & \textcolor{blue}{2.77} & \textcolor{blue}{3.48} & \textcolor{blue}{2.31} \\
 & Llama-3.2 3B & 0.9\% & 21.1\% & 38.8\% & \textcolor{red}{1.05} & \textcolor{red}{1.09} & \textcolor{red}{1.05} & \textcolor{blue}{2.64} & \textcolor{blue}{2.31} & \textcolor{blue}{2.13} \\
 & Qwen2-7B & 1.1\% & 0.2\% & 17.2\% & \textcolor{red}{1.06} & \textcolor{red}{1.16} & \textcolor{red}{1.18} & \textcolor{blue}{2.60} & \textcolor{blue}{2.96} & \textcolor{blue}{3.51} \\
 & Llama-3 8B & 0.8\% & 10.6\% & 18.0\% & \textcolor{red}{1.11} & \textcolor{red}{1.14} & \textcolor{red}{1.13} & \textcolor{blue}{2.57} & \textcolor{blue}{2.54} & \textcolor{blue}{2.80} \\
 & Mistral-NeMo 12B & 0.8\% & 10.6\% & 18.0\% & \textcolor{red}{1.11} & \textcolor{red}{1.14} & \textcolor{red}{1.13} & \textcolor{blue}{2.57} & \textcolor{blue}{2.54} & \textcolor{blue}{2.80} \\
\midrule
\multirow{5}{*}{OpenCodeInstruct} & Gemma-2 2B & 0.0\% & 0.0\% & 0.0\% & 1.00 & 1.00 & 1.00 & \textcolor{blue}{2.78} & \textcolor{blue}{2.76} & \textcolor{blue}{5.03} \\
 & Llama-3.2 3B & 0.1\% & 0.0\% & 0.0\% & 1.00 & 1.00 & 1.00 & \textcolor{blue}{2.24} & \textcolor{blue}{2.82} & \textcolor{blue}{4.89} \\
 & Qwen2-7B & 0.1\% & 0.4\% & 2.7\% & \textcolor{red}{1.02} & \textcolor{red}{1.01} & \textcolor{red}{1.02} & \textcolor{blue}{2.37} & \textcolor{blue}{3.62} & \textcolor{blue}{4.08} \\
 & Llama-3 8B & 0.1\% & 0.2\% & 1.0\% & \textcolor{red}{1.01} & \textcolor{red}{1.01} & 1.00 & \textcolor{blue}{2.24} & \textcolor{blue}{2.77} & \textcolor{blue}{4.77} \\
 & Mistral-NeMo 12B & 0.1\% & 0.2\% & 1.0\% & \textcolor{red}{1.01} & \textcolor{red}{1.01} & 1.00 & \textcolor{blue}{2.24} & \textcolor{blue}{2.77} & \textcolor{blue}{4.77} \\
\midrule
\multirow{5}{*}{ShareChat} & Gemma-2 2B & 0.1\% & 0.0\% & 0.1\% & 1.00 & 1.00 & 1.00 & \textcolor{blue}{2.23} & \textcolor{blue}{3.73} & \textcolor{blue}{3.31} \\
 & Llama-3.2 3B & 0.1\% & 0.0\% & 1.1\% & 1.00 & 1.00 & 1.00 & \textcolor{blue}{2.38} & \textcolor{blue}{3.65} & \textcolor{blue}{4.20} \\
 & Qwen2-7B & 0.0\% & 0.1\% & 1.9\% & \textcolor{red}{1.01} & \textcolor{red}{1.02} & \textcolor{red}{1.02} & \textcolor{blue}{2.82} & \textcolor{blue}{3.55} & \textcolor{blue}{3.94} \\
 & Llama-3 8B & 0.1\% & 0.0\% & 0.0\% & \textcolor{red}{1.01} & 1.00 & \textcolor{red}{1.01} & \textcolor{blue}{2.49} & \textcolor{blue}{3.65} & \textcolor{blue}{4.51} \\
 & Mistral-NeMo 12B & 0.1\% & 0.0\% & 0.0\% & \textcolor{red}{1.01} & 1.00 & \textcolor{red}{1.01} & \textcolor{blue}{2.49} & \textcolor{blue}{3.65} & \textcolor{blue}{4.51} \\
\bottomrule
\end{tabular}%
}
\end{table*}

\subsection{Serving-optimized Hybrid GC-SRAM Cache Organization}
\begin{figure}[h]
    \centering
    \includegraphics[width=\linewidth]{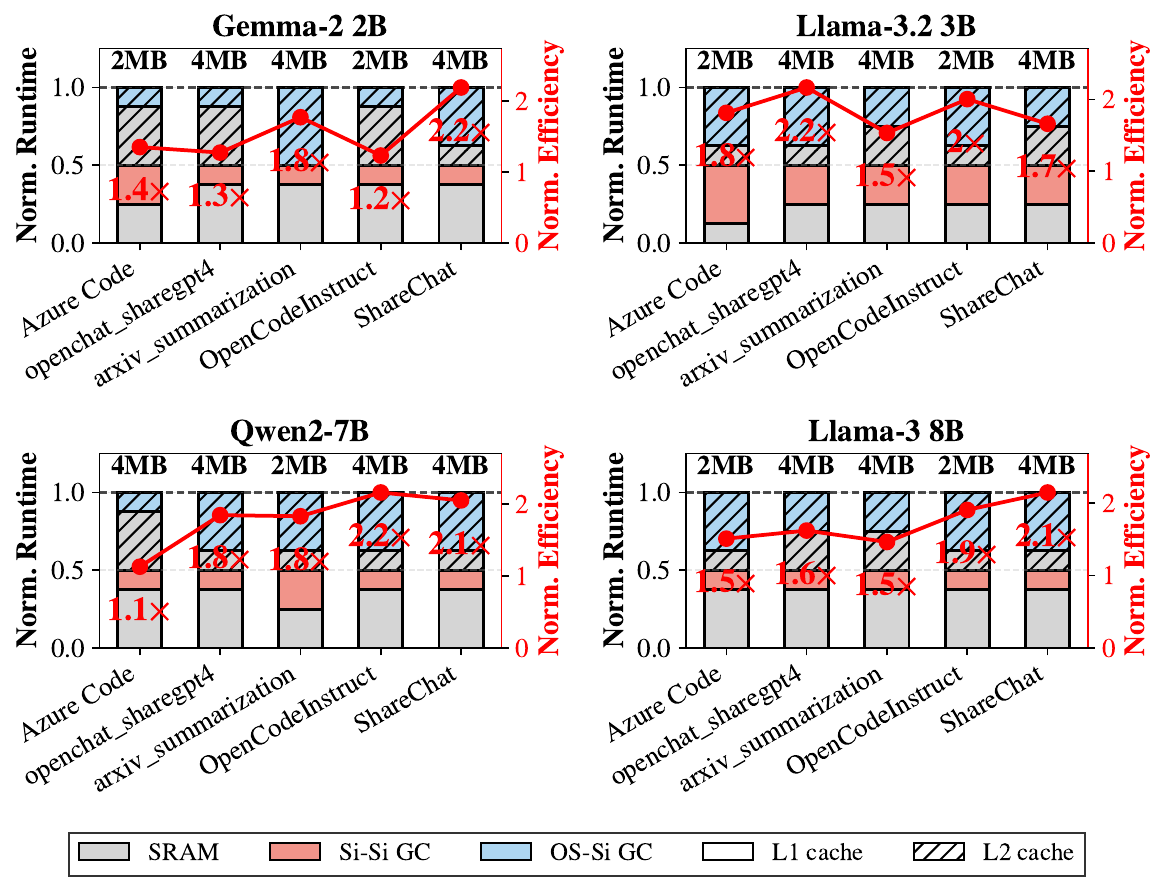}
    \caption{Hybrid cache hierarchy composition for minimizing total runtime for different workloads at batch size 32, showing normalized runtime ($\downarrow$ is better) and efficiency ($\uparrow$ is better) of best found hybrid cache design with respect to SRAM. Annotation denotes most optimal $L_2$ cache size found for hybrid cache.}
    \label{fig:dse_delay}
\end{figure}

\begin{figure}[h]
    \centering
    \includegraphics[width=\linewidth]{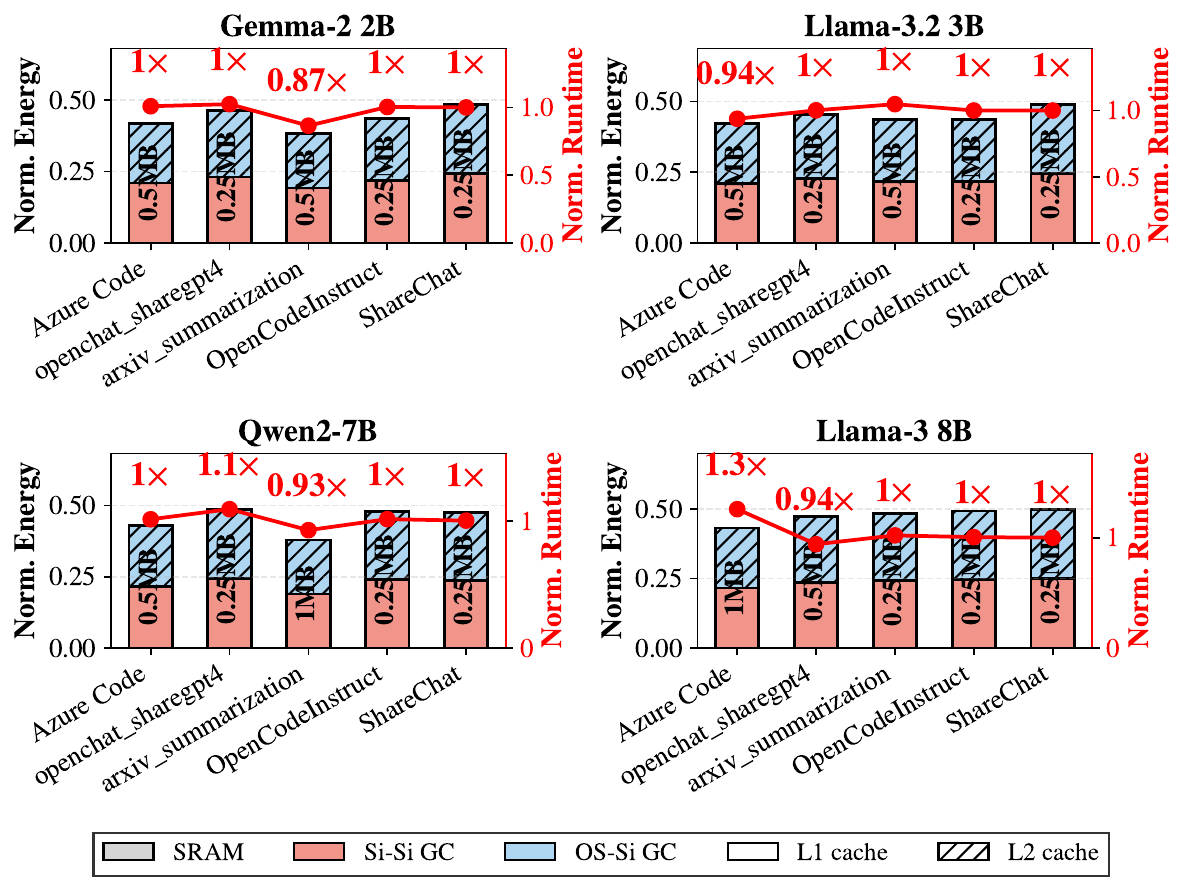}
    \caption{Hybrid cache hierarchy composition for minimizing energy for different workloads at batch size 32, showing normalized energy ($\downarrow$ is better) and runtime ($\downarrow$ is better) of best found hybrid cache design with respect to SRAM. Annotation denotes most optimal $L_2$ cache size found for hybrid cache.}
    \label{fig:dse_energy}
\end{figure}

\begin{figure}[h]
    \centering
    \includegraphics[width=\linewidth]{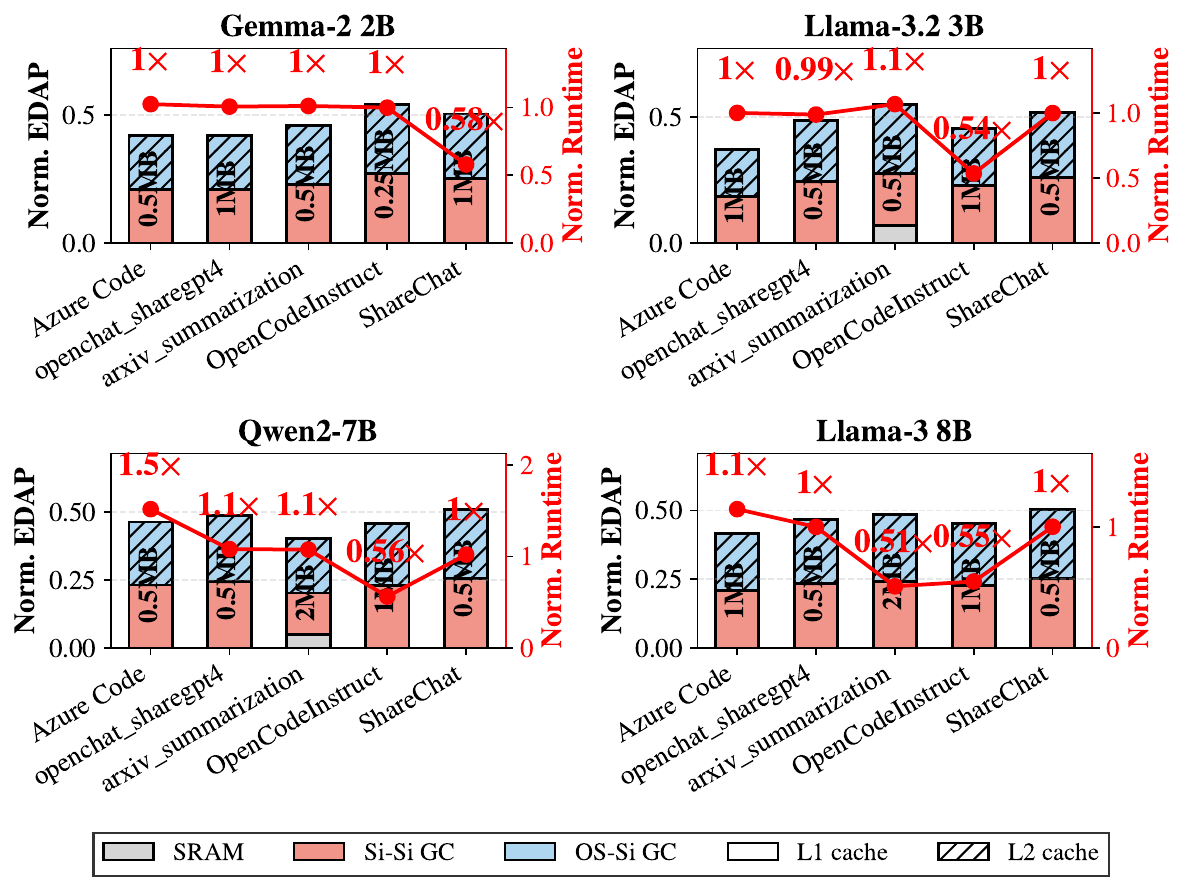}
    \caption{Hybrid cache hierarchy composition for minimizing EDAP for different workloads at batch size 32, showing normalized EDAP ($\downarrow$ is better) and runtime ($\downarrow$ is better) of best found hybrid cache design with respect to SRAM. Annotation denotes most optimal $L_2$ cache size found for hybrid cache.}
    \label{fig:dse_edap}
\end{figure}

In this experiment, we aim to identify the optimal hybrid GC-SRAM cache organization to optimize these three inference objectives separately: runtime, energy and energy-delay-area product (EDAP). We perform an extensive DSE over $f_{GC,L_1}, f_{GC,L_2}\in \{0.0, 0.25, 0.5, 0.75, 1.0\}$ and per-core $L_2$ cache size $\in \{256, 512, 1024, 2048, 4096\}$~kB across the different (workload, model) pairs. We use a batch size of 32, which exhibits the worst throughput degradation (Tables~\ref{tab:gc_frac_efficiency},~\ref{tab:ubuf_ablation}) and therefore demands the most nuanced hybrid design. For each objective, the \emph{Cost} function of the Mapper and Scheduler (Sec.~\ref{sec:sched}) is set to the target metric, and the best SRAM-only design ($f_{GC}=0$ for both $L_1$ and $L_2$) minimizing that metric serves as the baseline. Figs.~\ref{fig:dse_delay}--\ref{fig:dse_edap} report the resulting $L_1$/$L_2$ compositions with stack heights denoting the SRAM and GainCell fractions.

\noindent\textbf{Runtime:} Fig.~\ref{fig:dse_delay} shows the $L_1$ and $L_2$ cache compositions of the best hybrid GC-SRAM design, with bars indicating its normalized runtime with respect to the SRAM-only baseline. Clearly, for every workload, the best hybrid cache design matches the runtime of the best SRAM-only design while improving efficiency by up to $\mathbf{2.2\times}$ on decode-heavy and $\mathbf{1.8\times}$ on prefill-heavy workloads. This typically requires a larger SRAM fraction in $L_1$, which directly feeds the compute cores and must sustain high bandwidth to avoid stalls. This underscores the need for \ourwork{} to search the cache organization, delivering meaningful efficiency gains without sacrificing inference speed.\\

\noindent\textbf{Energy and EDAP:} Fig.\ref{fig:dse_energy} and \ref{fig:dse_edap} summarize our findings. Both the objectives are found to be minimized by a pure GainCell composition (\texttt{arxiv\_summarization} is an occasional exception for EDAP with hybrid GC-SRAM $L_1$ cache), yielding $\mathbf{\geq2\times}$ improvement over the best SRAM-only design. Notably, these designs also match or improve normalized runtime (\textcolor{red}{right y-axis} in Fig.\ref{fig:dse_energy},\ref{fig:dse_edap}), owing to GainCell-aware scheduling (Sec.~\ref{sec:hw_eff_result}, Table~\ref{tab:gc_frac_efficiency}). \textit{Hence, we conclude that for efficiency-centric serving, replacing SRAMs fully with Si-Si and OS-Si GCs at the $L_1$ and $L_2$ cache unlocks significant system-level efficiency over SRAM-based designs while achieving iso-throughput.}

\section{Conclusion}
We presented \emph{Let2Gain}, an end-to-end framework for evaluating 2T
GainCell memories as on-chip cache in 2.5D chiplet-based LLM accelerators. Our results show that GainCells' long-standing barriers to general-purpose caching can largely be overcome at the system level through GainCell-aware scheduling. LLM inference data lifetimes are short enough that nearly all tiles expire within the retention window, keeping refresh energy negligible at nominal cache sizes. We observe that replacing the SRAM with specific GainCells substantially lowers inference energy, boosts efficiency, and reduces peak temperature at slight throughput degradation occasionally. \ourwork{} further enables searching fine-grained hybrid GainCell-SRAM cache compositions that can match SRAM-only design's throughput at improved efficiency, allowing us to fully replace the SRAM stack. Future works can aim at hybrid GainCell-SRAM cache design for emerging spatial architectures such as 3D-DRAM and wafer-scale accelerators.

\section*{Acknowledgment}
The authors would like to thank Xiaochen Peng and
Kerem Akarvardar from TSMC Corporate Research, San Jose, CA 95134, USA for their valuable technical discussions and feedback on the manuscript. 

\noindent\textbf{Statement on Generative AI Usage:} The authors used Claude Opus-5.5 solely to correct grammar and
improve the readability of author-written text in this manuscript. The AI
tool was not used to generate research ideas, design or run experiments, produce or analyze data, or draw
conclusions. All AI-suggested edits were reviewed and approved by the
authors, who take full responsibility for the content of this article.

\bibliographystyle{IEEEtran}
\bibliography{IEEEabrv}
\end{document}